\documentclass[%
 aip,
 sd,%
 amsmath,amssymb,
 reprint,%
]{revtex4-1}

\usepackage{graphicx}
\usepackage{dcolumn}
\usepackage{bm}
\usepackage{changes}
\usepackage{siunitx}

\usepackage{comment}

\usepackage{subfig}
\usepackage[T1]{fontenc}

\usepackage[draft,nomargin,inline,author=]{fixme}
\fxsetup{theme=color}
\definecolor{fxwarning}{rgb}{0.8,0.0000,0.0000}

\begin{document}

\preprint{AIP/123-QED}

\title{A Laser Spiking Neuron in a Photonic Integrated Circuit}

\author{Mitchell A. Nahmias*}
\email{mitch@luminouscomputing.com}
\author{Hsuan-Tung Peng*}
\author{Thomas Ferreira de Lima*}
\author{Chaoran Huang}
\author{Alexander N. Tait}
\author{Bhavin J. Shastri}
\author{Paul R. Prucnal}

\thanks{*These authors have contributed equally to this work.}

\affiliation{
Electrical Engineering Department, Princeton University\\41 Olden St, Princeton, NJ 08540, USA
}%

\date{\today}

\begin{abstract}
There has been a recent surge of interest in the implementation of linear operations such as matrix multipications using photonic integrated circuit technology. However, these approaches require an efficient and flexible way to perform nonlinear operations in the photonic domain. We have fabricated an optoelectronic nonlinear device---a \emph{laser neuron}---that uses excitable laser dynamics to achieve biologically-inspired spiking behavior.
We demonstrate functionality with simultaneous excitation, inhibition, and summation across multiple wavelengths. We also demonstrate cascadability and compatibility with a wavelength multiplexing protocol, both essential for larger scale system integration.
Laser neurons represent an important class of optoelectronic nonlinear processors that can complement both the enormous bandwidth density and energy efficiency of photonic computing operations.
\end{abstract}


\maketitle

\begin{figure*}
\centering
\includegraphics[width=1\linewidth]{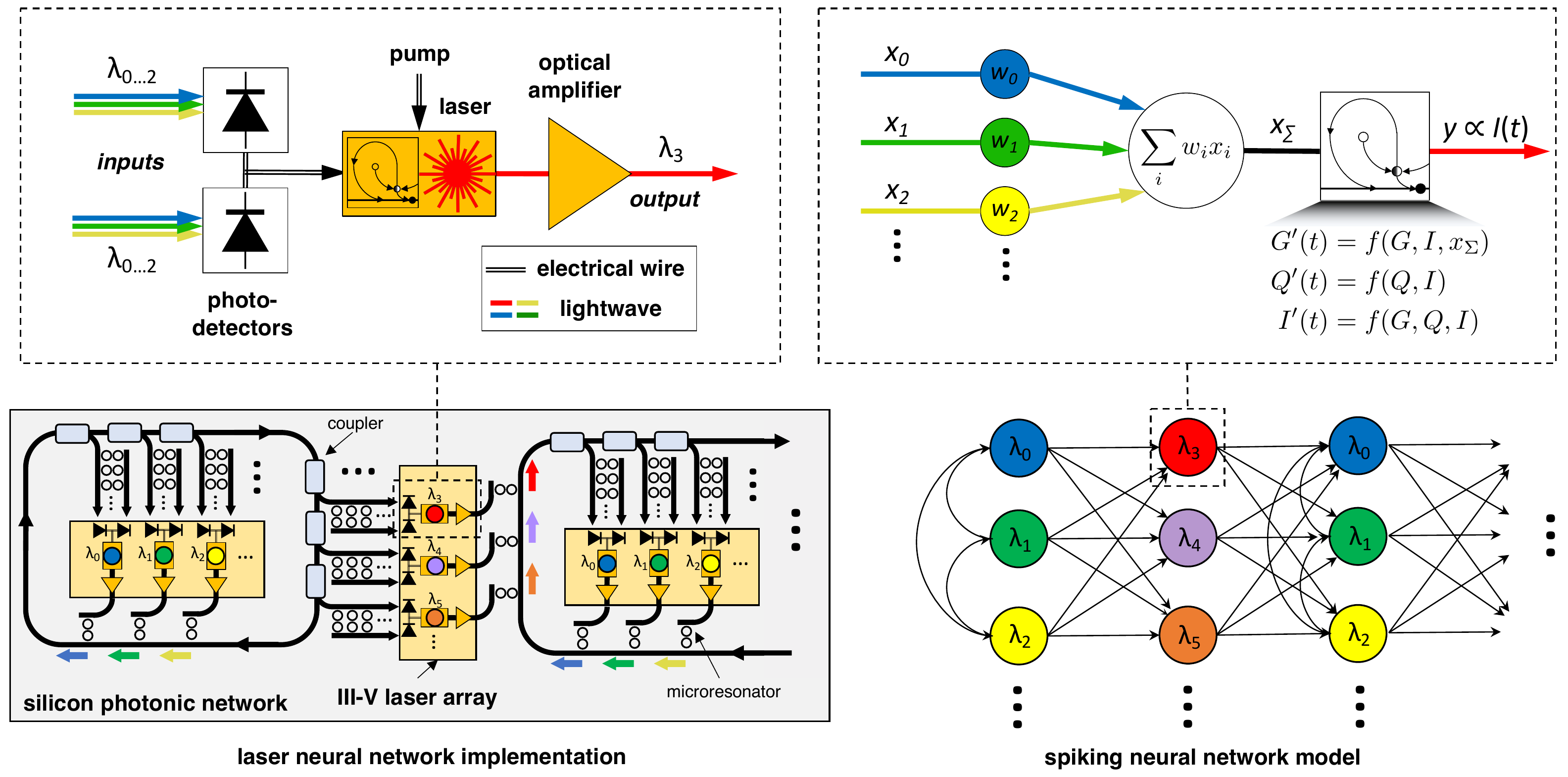}
\caption{Schematic of a spiking neural network implemented in a photonic integrated circuit (PIC). Networks can be instantiated using III-V laser arrays bonded to silicon photonic chips (bottom left), which include passive couplers and microresonators. Both recurrent and feedforward network topologies are possible using the B\&W framework, which assigns a unique wavelength $\lambda_i$ to each laser neuron (bottom right). The processing node itself consists of a pair of balanced photodetectors connected to a laser followed by a semiconductor optical amplifier, emulating a biological spiking neuron (top left). The photodetectors perform a summation operation, while the gain $G$, loss $Q$, and cavity intensity $I$ interact to generation excitable dynamics (top right). \label{fig:concept}}
\end{figure*}


High performance computing has experienced accelerating growth in the last decade, driven largely by the rapid expansion of machine learning applications.
For example, deep learning training
is doubling at a rate at 3.5 months, far outpacing Moore's law of performance doubling every 18 months~\cite{openai-compute}. This gap in supply and demand is exacerbated by the increasing difficulty of continuing Moore's law in hardware: since electronic devices
are reaching feature size limits and are no longer subject to Dennard's law~\cite{Esmaeilzadeh2012}, they require more exotic geometries and material platforms to sustain their past exponential growth in performance~\cite{irds2017}.

These limitations, together with the vast computing requirements of artificial intelligence, have motivated the development of application specific integrated circuits (ASICs) for deep learning, 
a notable example of which is Google's tensor processing unit (TPU)~\cite{Jouppi:2017aa}.
More exotic approaches involving non-volatile, co-located memory~\cite{Yu:2018aa,Ielmini:2018aa,Burr:2017} including phase-change analog~\cite{Ambrogio:2018aa}
or memristors~\cite{Jo:2010,Prezioso:2015aa,Bayat:2018aa}
promise orders of magnitude increases in efficiency and processing density. However, electronic approaches must grapple with two significant sources of energy consumption: data movement---especially between the memory and processor---and capacity for compute (i.e., operations per second), largely dominated by linear operations such as matrix multiplications.


Photonics has been well studied for its potential to address both bottlenecks (see Ref.~\cite{miller2009device,Psaltis:88}). Electronic data movement involves capacitively charging and discharging metal interconnects, with energy consumption that is roughly proportional to the length of each wire. In contrast, although photonic channels require energy for E/O or O/E conversion, it is no longer the critical path in transceivers~\cite{Georgas:2011aa}, and the energy consumption of each link scales nearly independently of its length.
Current photonic systems are competitive with on-chip electronic interconnects (<1 pJ/bit), and will increase in efficiency as optoelectronic devices
see continued improvements~\cite{Miller:17}.

Deep learning compute primarily involves matrix-vector multiplications, which are composed of multiply-accumulate (MAC) operations: a single operation consists of $a \leftarrow a + w \times x$ with accumulation variable $a$, signal $x$, and weight $w$.
For these operations, photonic components exhibit major advantages over digital electronics in energy, speed, and computational power. 
First, as noted by Ref.~\cite{10.3389/fnins.2015.00484}, passive analog energy consumption is not necessarily proportional to the number of operations being performed. As an example, for a matrix operation with $M$-sized vector inputs and outputs, the number of computations is proportional to $M^2$, but the signal generation cost is proportional to the number of channels $M$. This property also extends to photonic systems~\cite{Shen:2017aa}. Secondly, photonic components can operate at much higher speeds (>\SI{5}{GHz}); they are not limited by thermal dissipation, clock distribution, and interconnect jitter. Third, digital MAC operations---typically implemented via adders and multipliers---requires thousands of transistors, whereas photonic MAC operations only require one (or several) passive photonic devices to accomplish the same functionality. This simplicity, together with a higher clock rate, allow on-chip photonics to exhibit higher processing densities than state-of-the-art digital electronic matrix multipliers, despite the large sizes of photonic devices~\cite{PrucnalBook}.

However, implementing nonlinear operations or interfacing with stored digital data requires high speed analog-to-digital conversion, which can consume a significant amount of energy~\cite{Walden:1999aa,Walden:2008}. Instead, photonic nonlinearities can reduce the number of conversion steps by implementing many processing layers in the photonic domain. However, current approaches, which include resonator-enhanced optical nonlinearities~\cite{Tait:2013,Notomi:07,Nozaki:2010aa,Van:2002aa} or optoelectronic nonlinearites~\cite{Ren:2011aa,Kauranen:2012aa,takahashi1996ultrafast}, require either exotic materials or large threshold powers. They also have difficulty exhibiting complex nonlinear behaviors such as spiking. To this end, a number of approaches have explored approaches to emulate spiking functionality~\cite{Prucnal:16,Feldmann:2019aa}, but many of them require specialized devices which are incompatible with emerging standards in the photonic integrated circuit (PIC) industry.

In this paper, we demonstrate that a \emph{laser neuron}---consisting of a balanced photodetector pair that directly modulates a laser~\cite{Peng:2018aa,Peng:2019aa}---can
emulate a Leaky Integrate-and-Fire (LIF) neuron---the most widely used model in computational neuroscience---across many simultaneous wavelength channels in a standard PIC platform. 
In contrast to their microelectronic counterparts, laser neurons can process data at high speeds $( > \mathrm{GHz})$ while dissipating relatively little energy during data movement.
We experimentally demonstrate a variety of critical functions, characterize speed and energy consumption, and discuss strategies for implementing units into larger-scale networks.

\section{Laser Neuron Architecture}

\begin{figure*}
\centering
\includegraphics[width=1\linewidth]{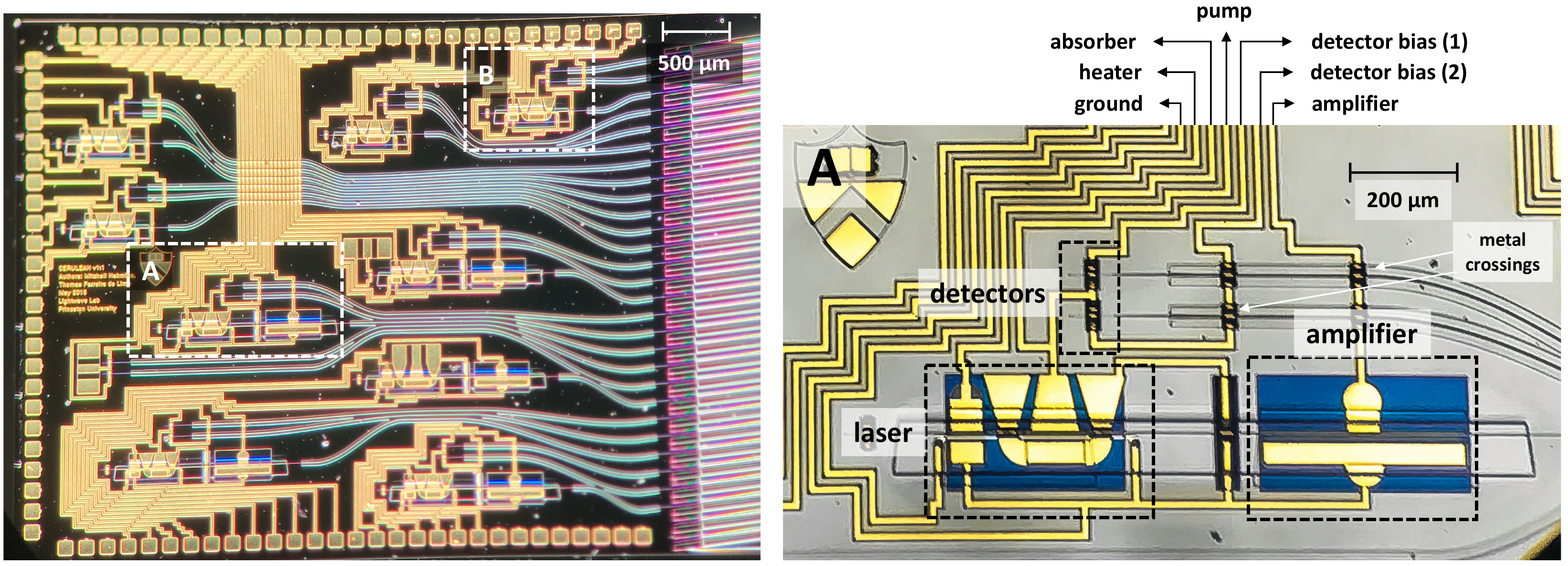}
\caption{(Left) Photograph of III-V PIC with an array of laser neuron test structures, fabricated at the Heinrich-Hertz Institute, with models that include (type {\bf A}) or do not include (type {\bf B}) an output amplifier. (Right) Micrograph of single laser neuron unit, consisting of a balanced photodetector (BPD) pair, a two-section distributed feedback (DFB) laser, and a semiconductor optical amplifier (SOA) on the output port. \label{fig:chip}}
\end{figure*}

\subsection{Model}

Each laser neuron models the behavior of a simple spiking neuron using optical pulses to code information (for a further discussion on spiking, see the Supplementary materials).
As shown in Figure~\ref{fig:concept}, a single processing unit consists of a pair of photodetectors directly wired to the input terminal of a laser, followed by an amplifier. Inputs of multiple wavelengths $\lambda_0, \lambda_1, \dots \lambda_M$---in which the intensites $x_i$ are weighted by $w_i$ using a passive silicon network external to each processing unit---are incident on a pair of balanced photodetectors. Excited carriers relax and sum together the $M$ input signals, resulting in a current signal proportional to $\sum w_i x_i$. The resulting push-pull current travels into a laser biased just below threshold. The input acts as a perturbation to the laser's internal dynamical system $\dot{s} = f(s)$, and with enough positive inputs, the laser can \emph{excite} and fire an optical pulse as the output $y(t)$.


The laser's dynamical system is represented via the interactions between a gain medium, an absorbing medium, and the light within the cavity. This system performs several nonlinear processing functions on the input data, including integration, thresholding, and time discretization (i.,e., refractoriness) via a mechanism called \emph{excitability}. A simplified, undimensionalized version of the equations governing this system can be represented by~\cite{Dubbeldam1999}:
\begin{align}
\label{eq:Yamada}
  \dot{G(t)}&=\gamma_G\left[A-G(t)-G(t)I(t)\right]\\
  \dot{Q(t)}&=\gamma_Q\left[B-Q(t)-aQ(t)I(t)\right]\\
  \dot{I(t)}&=\gamma_I\left[G(t)-Q(t)-1\right]I(t)+\epsilon f(G)
\end{align}
for gain variable $G(t)$, absorber variable $Q(t)$, cavity intensity $I(t)$, and parameters ($A$,$B$,$a$,$\gamma_G$,$\gamma_Q$,$\gamma_I$,$\epsilon$).
As discovered in Ref.~\cite{Nahmias2013}, under certain conditions, these equations simplify to a model of a Leaky Integrate-and-Fire (LIF) neuron, a popular spiking model in computational neuroscience~\cite{Koch:1998}:
\begin{align}
\dot{G(t)}&=-\gamma_G(G(t)-A)+\theta(t);\\
  &\text{if $G(t)>G_{\mathrm{thresh}}$ then}\\\nonumber
  &\text{release a pulse, and set $G(t)\rightarrow G_{\mathrm{reset}}$.}
\end{align}
Together, a balanced photodetector, laser and amplifier can emulate the basic functions of a spiking neuron. In principle, networks of spiking neurons can perform any algorithm or simulate any nonlinear dynamical system~\cite{maass2001pulsed}.

\subsection{Networking}

\begin{figure}
\centering
\includegraphics[width=.9\linewidth]{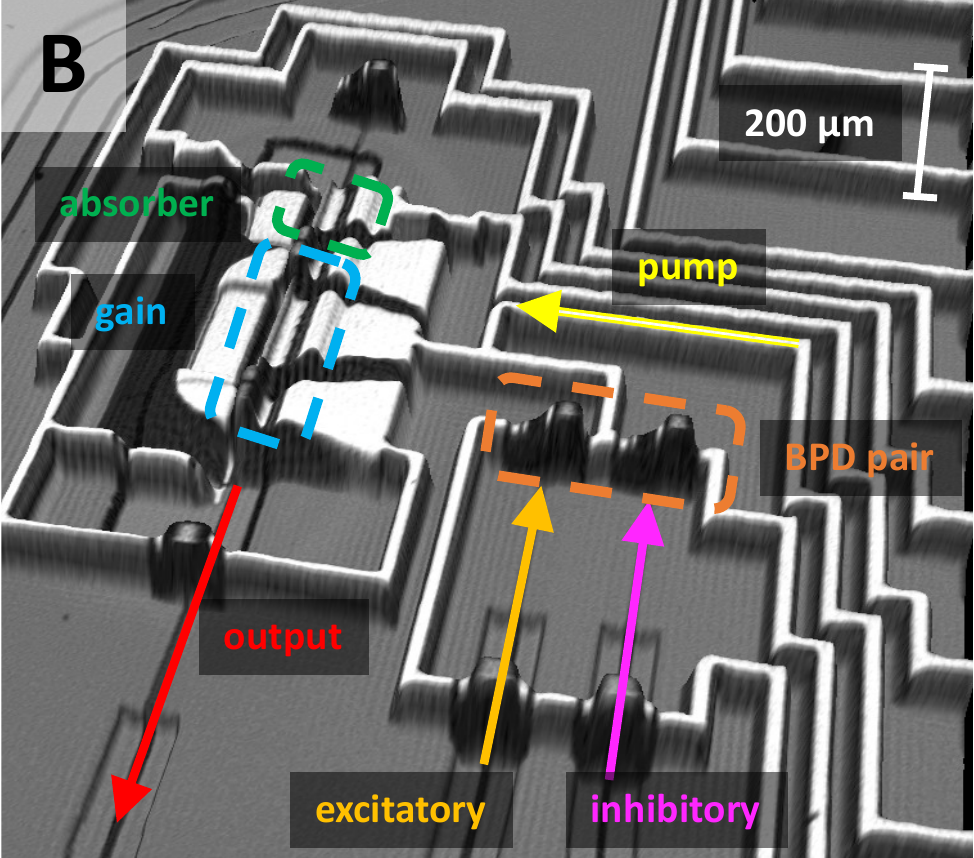}
\caption{Micrograph of a single laser neuron containing no amplifier on the output port (type {\bf B}). Inputs to photodetector pair either contribution positively (i.e., are excitatory) or negatively (i.e., are inhibitory). Biasing the absorber at different voltages alters properties of the optoelectronic nonlinearity, discussed in Ref.~\cite{Peng:2018aa}. \label{fig:chipzoom}}
\end{figure}

Laser neurons are designed to be compatible with Broadcast-and-Weight (B\&W), a reconfigurable optical neural networking method proposed in Ref.~\cite{Tait:JLT:2014}. The B\&W protocol assigns each laser neuron a unique wavelength $\lambda_i$. Wavelength division multiplexing (WDM) allows for the aggregation of these signals along common bus waveguides, which distribute the signals with unique transmission profiles to each processing node. Tunable filter banks adjust the strength of each connection, or weight\cite{Tait:16anal,Tait:18}. The resulting weighted signals sum together via the balanced photodetectors (BPDs) driving each laser. This system can implement both negative and positive weights ($w_{ij} \in [-1, 1]$), allowing for fully reconfigurable neural network models. Coupling to a waveguides in a loop topology allows for recurrent connections, while coupling from one waveguide to another allows for feedforward connections, as illustrated in Fig~\ref{fig:concept}.

In contrast to several other networking frameworks (i.e., coherent matrix multiplication~\cite{Shen:2017aa} or optical reservoirs~\cite{Appeltant:2011aa,Paquot:2012,Brunner:2013aa,Vandoorne:2014aa}), B\&W can implement both feedforward and recurrent connections with full tunability. An example of a network including both types is illustrated in Fig.~\ref{fig:concept} (bottom left), in which two recurrent groups of neurons communicate via a feedforward set of laser neurons. The B\&W protocol can also realize more complex and interesting topologies such as hierarchical or small-world networks, as discussed in Ref.~\cite{Tait:JLT:2014,Peng:2018aa}.

\subsection{Fabrication}

Each laser neuron consists of III-V photonic devices that can be found in most standard process design kits (PDKs) common to large-scale foundry models:
a distributed feedback laser (DFB), a balanced photodetector (BPD) pair, and a semiconductor optical amplifier (SOA). Lithographically-defined metal wires connect the components together in a way that allows for direct interactions between the detectors and the laser. The DFB lasers are composed of electrically pumped multi-quantum wells (MQW) with emission near $\SI{\sim1550}{nm}$ embedded in a ridge-waveguide structure.
The laser includes both a primary gain section as well as a smaller absorber section, with lengths $L_G = \SI{125.0}{\micro\meter}$ and $L_Q = \SI{75.0}{\micro\meter}$, respectively. An etched, intracavity electrical isolation section of length $L_\mathrm{iso} = \SI{75.0}{\micro\meter}$ divides the two sections, and a small absorber placed on the non-emitting port of each laser reduces back reflections. The SOA also includes an active MQW structure, with a length set to $L_\mathrm{SOA} = \SI{400.0}{\micro\meter}$. Device layouts were generated in collaboration with the Fraunhofer Institute for Telecommunications, at the Heinrich Hertz Institute (HHI) as part of the Joint European Platform for Photonic Integration of Components and Circuits (JePPiX) consortium.

\begin{figure}
\centering
\subfloat[Multi-Channel Spiking Demonstration]{
\includegraphics[width=.95\linewidth]{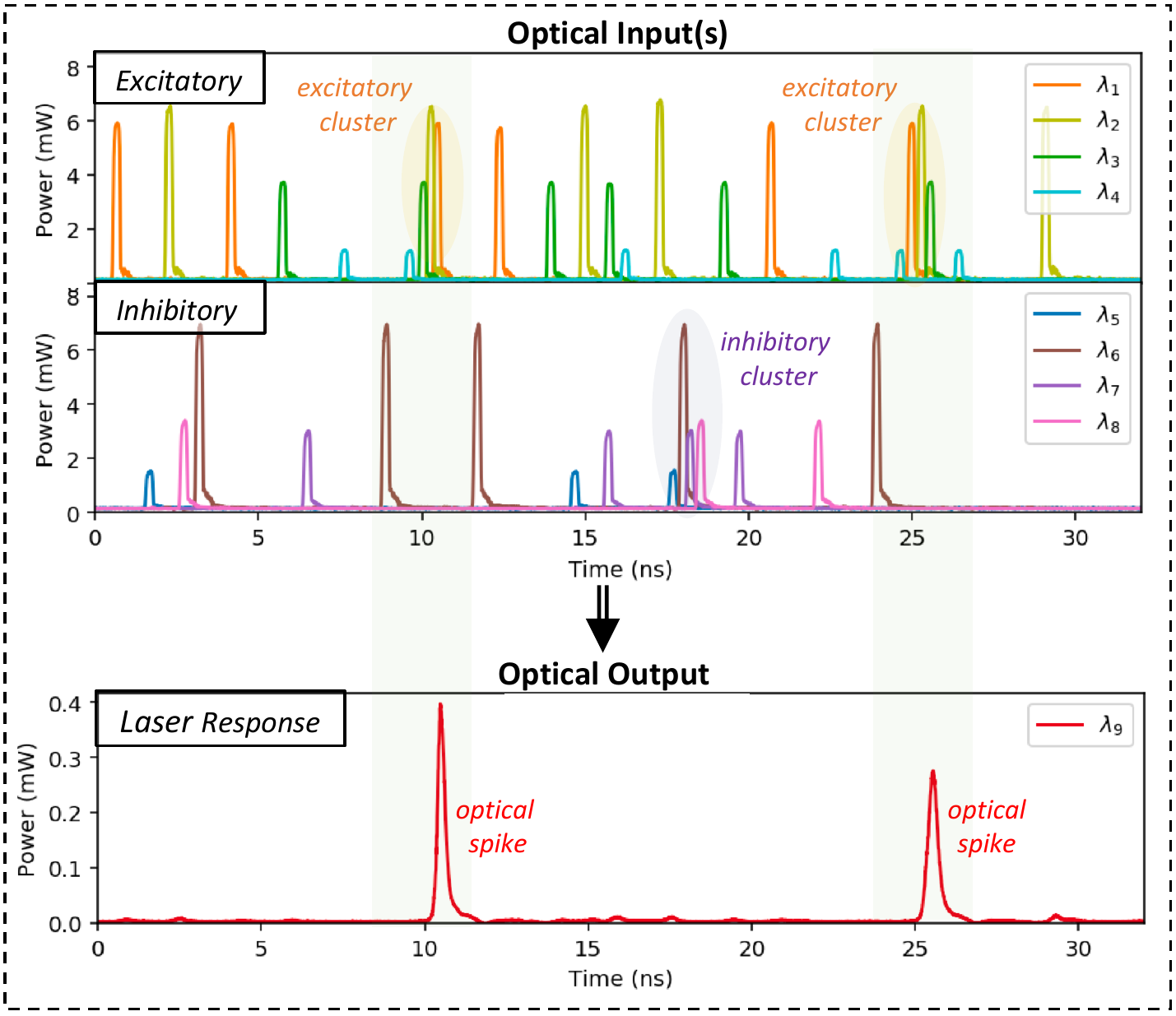}} \\
\subfloat[Multi-Channel Inhibitory Suppression]{
\includegraphics[width=.95\linewidth]{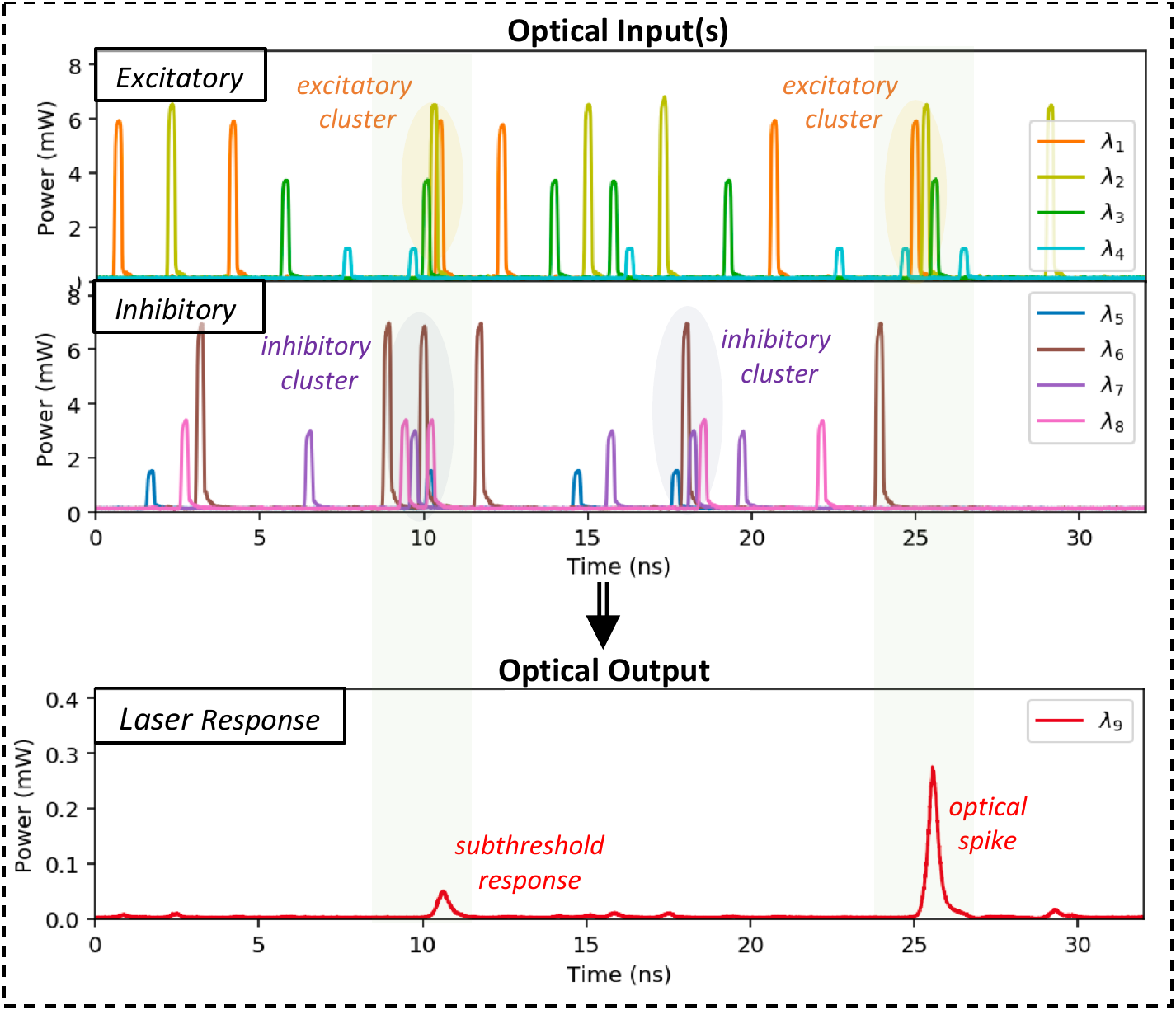}}
\caption{Experimental traces of multi-channel spiking behavior in a laser neuron with no output amplifier (type {\bf B} in Fig.~\ref{fig:chip}) tested over eight wavelengths $\lambda_{1-8}$.
(a) Clusters of excitatory pulses (top) incident on the excitatory detector can trigger the generation of optical pulses by the laser (bottom) if in close temporal proximity. (b) Clusters of inhibitory pulses (middle) have a negative effect, and can cancel out the positive effect of excitatory pulses, resulting in a subthreshold response (bottom).
\label{fig:traces}}
\end{figure}

\begin{figure}
\centering
\includegraphics[width=1\linewidth]{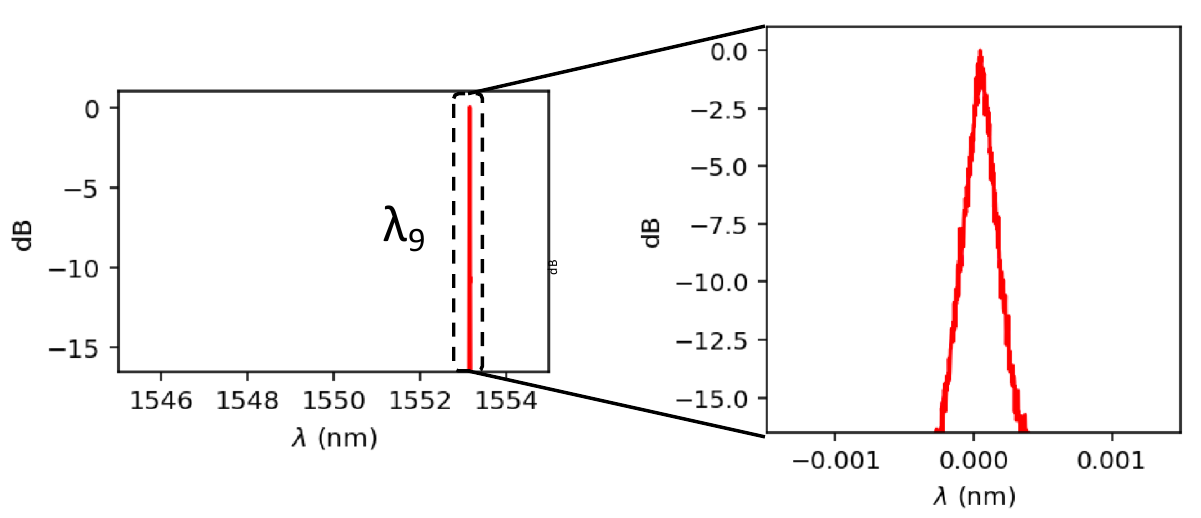}
\caption{Measured output spectrum of a type {\bf B} laser neuron during operation (peak normalized to \SI{0}{dB}). Light is generated by a distributed feedback laser and outputs with a narrow linewidth $\Delta \lambda <\SI{.001}{nm}$ (right).}
\label{fig:spectrum}
\end{figure}

\section{Results}

\subsection{Multi-Wavelength Functionality}

To utilize the dense interconnectivity possible in the B\&W protocol, a laser spiking neuron $j$ must be able to receive $M$ intensity signals with unique wavelengths $\lambda_1, \lambda_2, \dots \lambda_{M}$, and emit a single wavelength $\lambda_j$. We experimentally demonstrate this functionality together with spiking dynamics, measuring the nonlinear response to both excitatory (positive) and inhibitory (negative) pulses (Fig.~\ref{fig:traces}), and measure the output spectrum for above-threshold signals (Fig.~\ref{fig:spectrum}). For this experiment, we used a laser neuron unit without amplifier (unit B, as illustrated in Fig.~\ref{fig:chipzoom}).

For simplicity, the experimental demonstration used a total of eight wavelength channels with independent spiking signals: four inputs incident on each photodetector. A topographical micrograph of the device is shown in Fig.~\ref{fig:chipzoom}.
The laser current is biased just below the lasing threshold (\SI{11}{mA}) to initiate the system in a state of excitability. The excitatory photodetector is reversed biased at \SI{3.6}{V}, whereas the inhibitory photodetector is reversed biased at \SI{1.43}{V}. As shown in Fig~\ref{fig:traces}(a), the laser neuron only responds with an output pulse if a cluster of excitatory pulses arrives closely spaced in time. This demonstrates several key attributes: summation across multiple wavelengths inputs $x_i |_{\lambda_i}$, the ability to integrate pulse activity across some integration time interval $T_\mathrm{int}$, and the ability to make a binarized (0,1) threshold decision based on input spike activity.

A BPD pair allows for the implementation of both positive and negative weights. Fig~\ref{fig:traces}(b) shows that inhibitory pulses can oppose the activity excitatory pulses: generating a cluster of inhibitory pulses that coincide with the first excitatory cluster results in a cancellation of the output pulse (i.e., via a \emph{negative weighting} of the inhibitory signals that opposes the \emph{positive weighting} of the excitatory signals). Another important condition is a well-defined output wavelength, critical for wavelength identification and filtering in the B\&W protocol. Fig~\ref{fig:spectrum} shows the output spectrum of the laser modulated with excitatory pulses above threshold: it outputs with a stable and narrow linewidth $\Delta \lambda < \SI{.001}{nm}$.

Although just eight channels were demonstrated here, the B\&W protocol allows for flexible channel scaling through the addition of more wavelengths and laser processing nodes.
In B\&W networks, two primary limiting factors include the finesse $\mathcal{F}$ of the passive filters in the network~\cite{Tait:16anal}, and the gain bandwidth of the lasers. If high finesse, miniaturized resonators~\cite{Timurdogan:2014aa} are combined with a standard III-V laser gain spectrum~\cite{Arakawa:1986aa} covering the optical C-band, $M$ can reach on the order of several hundred channels.

This allows us to characterize the potential \emph{processing speed} of each laser as a function of the number of wavelength channels $M$.
With refractory period $\Delta T$, a single laser neuron can make a spike or no-spike decision across $M$ inputs every $\Delta T$. The number of MACs per second---the speed of each processor---is therefore $S = M/\Delta T$. With a refractory period $\Delta T \sim \SI{200}{ps}$~\cite{Peng:IPC} and $M \sim 200$ (assuming < 3 dB power penalty, see Ref.~\cite{Tait:18,Xu:2008aa}), we arrive at \SI{1e12}{MACs/s}, or \SI{1}{TMAC/s}. This speed is quite enormous for a single device, exceeding the total processing capacity of many microelectronic processors.
Note that since the speed per node is $\propto M$,
it is dependent on channel scalability: higher processing capacity requires a fully-scaled processing system as depicted in Fig.~\ref{fig:concept} with several hundred wavelength channels per broadcast waveguide.



\subsection{Cascadability}
\label{sec:cascadability}

\begin{figure*}
\centering
\includegraphics[width=1\linewidth]{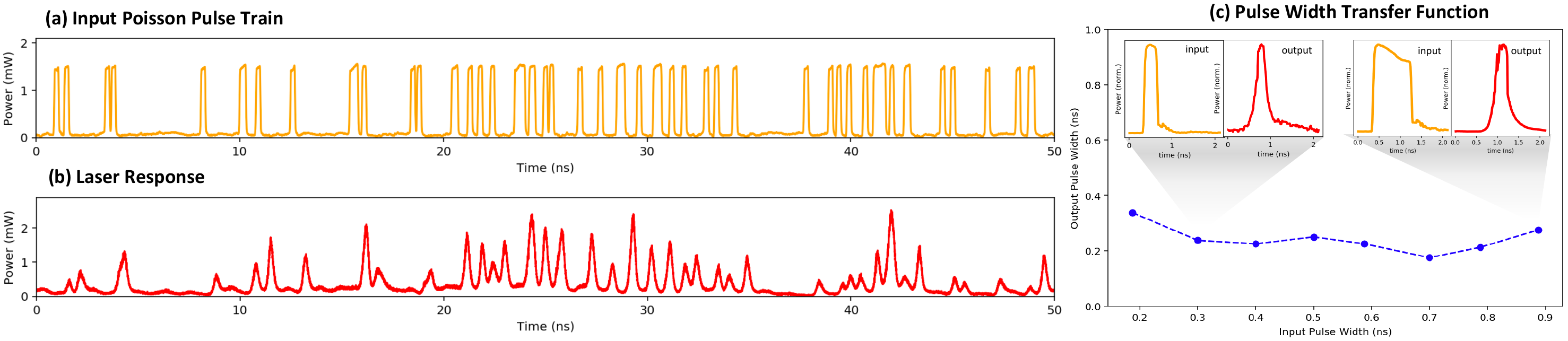}
\caption{(a,b) Input and output traces fulfilling the closed-loop gain condition, tested with a laser neuron with an output amplifier (type {\bf A}). Inputs were generated using a Poisson point process with $\mu_p = \SI{1}{GHz}$ and pulse widths $\tau_p = \SI{0.2}{ns}$. Output was set with SOA current $I_\mathrm{SOA} = \SI{105}{mA}$ s.t. $\overline{P_\mathrm{out}} \approx \overline{P_\mathrm{in}}$. (c) Stabilization of pulse width (measured as full-width half maximum), demonstrating temporal pulse regeneration capabilities. Several traces at an input at a width of $\tau_\mathrm{in} = \SI{.3}{ns}$ and $\tau_\mathrm{in} = \SI{.9}{ns}$ are also shone. \label{fig:cascade}}
\end{figure*}

An important condition for larger networks is that the nonlinear processors are \emph{cascadable}. As discussed in Ref.~\cite{Peng:2018aa}, signals must be able to propagate through a network without degradation.
This divides into both \emph{gain} cascadability---the ability to drive the next stage of neurons with enough energy---and \emph{signal} cascadability, the fidelity of the information encoding from one stage to another. In this section, we experimentally demonstrate that a laser spiking neuron can meet a number of cascadability conditions in both domains, and characterize its performance and energy consumption during operation.

We first show that a laser neuron with an amplifier (type {\bf A} in Fig.~\ref{fig:chip}) can meet the closed-loop gain condition with fixed point precision (see Sec.~\ref{sec:precision} for a discussion on precision).
To simulate peak processing conditions, we generate a dense, random stream of excitatory inputs spikes and measure the laser output response. We assume a Poisson point process (a common assumption in spiking signals~\cite{Gabbiani:1998}). The probability distribution of $N(t)$, where $N(t)$ is the number of spikes that occur on the interval $[0,t]$ is given by:
\begin{align}
P \{ N(t) = n \} = \frac{(\mu_p t )^n}{n!} e^{-\mu_p t}
\end{align}
We set $\mu_p = \SI{1}{GHz}$ and defined the input pulse width as $\tau_p = \SI{0.2}{ns}$ over a repeated time interval \SI{50}{ns} (see Sec.~\ref{sec:experimental} for more discussion on experimental conditions). Meeting the closed loop gain condition requires that the optical output power exceeds the input ($P_\mathrm{out} > P_\mathrm{in}$). 
We adjusted the SOA input current until the output power exceeds the input by about \SI{\sim3}{dB} to account for the projected coupling and microresonator insertion losses in the system.
This occured for an SOA current of $I_\mathrm{SOA} > \SI{105}{mA}$ at \SI{2.5}{V}. We show time traces of both the input and output in Fig.~\ref{fig:cascade} for this condition.

Secondly, we confirmed that each neuron has nonlinear pulse regeneration capabilities. As discussed in Ref.~\cite{Peng:2018aa}, to assure that spikes remain binarized over many stages, nonlinear processors must regenerate pulses as the are incident on each device. We demonstrated pulse width compression and stability (Fig.~\ref{fig:cascade}, right): pulses stay approximately constant a full-width half maximum (FWHM) width $\tau_p$ of \SIrange{0.2}{0.3}{ns}. This assures that spikes do not lose their timing characteristics as they propagate forward, maintaining a temporal precision of $\sim\tau_p$. The results indicate more than just a simple nonlinearity---input activity (even a square pulse, as shown in Fig.~\ref{fig:cascade}) manifests in the output as characteristic pulse with a fairly stable FWHM.

Based on these measurements, we can calculate the energy consumption of each laser processor.
The vast majority of dissipation occurs in the amplifier, consuming approximately $P_n = \SI{0.26}{W}$ per node. With a processing speed of $S_n = \SI{1}{TMACs/s}$, this amounts to $P_n/S_n = \SI{\sim260}{fJ}$ per MAC. This is the range of current deep learning hardware (i.e., see comparison in Ref.~\cite{Peng:2018aa}), and depends on a large channel number $M \sim 200$ to realize its advantages. Nonetheless, it is far from the most efficient processing model possible.
For example, efficient directly-driven lasers (i.e., lower threshold models~\cite{Takeda:2013aa,Wu:2015aa}) negate the need for an amplifier. Alternatively, to stay compatible with emerging PIC standards, transimpedance amplifiers in platforms with co-integration between electronics and photonics can provide efficient electrical gain between detectors and lasers (many examples of which are provided in Ref.~\cite{Jeong:2017}).




\section{Conclusion}

We have demonstrated that a laser neuron, fabricated in a photonic integrated circuit platform, can function as a processing node in a larger scale spiking neural network.
Laser neurons communicate photonically, sidestepping many of the costs associated with both data movement and the implementation of linear operations in electronics.
This leads to the potential for much higher speeds and energy efficiencies compared to neuromorphic electronic processors.
We experimentally validated LIF neuron model functionality across multiple wavelength channels,
including the ability to integrate multiple signals together across time, accept both positive (excitatory) and negative (inhibitory) inputs, and make a binary (0,1) spike classification based on pulsed activity. We verified its compatibility with the B\&W protocol, assuring that it can utilize the full bandwidth density available to optical waveguides for connectivity.
We also demonstrated cascadability, both in the laser neuron's ability to sustain and amplify signals and its ability to maintain the integrity of pulsed signals from one layer to another.

Our calculated speed and energy efficiencies---\SI{1}{TMAC/s} per neuron and \SI{260}{fJ/MAC}, respectively---exceed current microelectronic performance figures, particularly in speed.
Further developments in optoelectronic devices~\cite{Miller:17}, co-integration between photonic and electronic platforms~\cite{Sun:2015aa,Atabaki:2018aa}, or the utilization of novel materials such as graphene~\cite{Shastri:2016aa} provide ample avenues for further exploration. These techniques 
would realize the potential 3-5 orders-of-magnitude improvements~\cite{Peng:2018aa} that neuromorphic photonic computing has to offer.

\section{Supplementary Materials}

\subsection{Precision}

\label{sec:precision}

The precision of each computation, bounded by noise and device variations, limits information capacity.
We can define precision with respect to multiply-and-accumulate (MAC) operations: each laser neuron computes a dot product $\vec{w} \cdot \vec{x}$ of vector length $M$ with $k$ bits of precision, giving a total of $k \times M$ bits being computed over $M$ MAC operations.
To avoid processing degradation, cascadability requires that signals with $k$ bits of precision stay above some set threshold $k > k_T$. In the case of spiking, the amplitude is binary (1 bit of precision), while the spike times should remain analog ($\log_2(t_q/\tau_p)$ bits of precision for time between spikes $t_q$ for spike $q$ and pulse width $\tau_p$). Therefore, laser processors must assure that outputs remain spatially coherent, while preventing a reduction in analog temporal precision by keeping $\tau_p$ below a threshold value $\tau_p < \tau_{p(T)}$. The lack of this condition can eventually cause pulses to widen, and degrade as they propagate through a network.


MAC operations can either be floating point---in which the quantization threshold is proportional to the output amplitude, as seen in most digital processors---or they can be \emph{fixed point}, in which the quantization threshold is set independently of the output amplitude. AI researchers have shown that fixed point matrix multiplication can work just as well for deep learning models, even for training~\cite{Koster:2017}. Inference, in particular, does not require high resolution: typically several bits of precision can achieve near state-of-the-art performance\cite{Hubara:2016,Courbariaux:2015}.
B\&W weight networks best approximate fixed point linear operations, since precision is typically bounded by some physical noise threshold by the signal or reciever. In fixed point arithmetic, a minimum power resolution threshold $P_T$ is set at the detector, and the number of bits of precision for signal $P$ is equal to $N_b = \log_2(P/P_T)$. To prevent degradation, the total signal amplitude $\sum_i y_i$ from one stage to another must be conserved. Divided individually, a laser neuron must, on \emph{average}, provide enough gain to compensate for node-to-node losses.
In a fixed point framework, the total power required to meet this condition is proportional to the number of processors, $N$, not the number of connections $N M$, an advantage that arises from the non-dissipative nature of passive analog operations~\cite{Shen:2017aa}. If the B\&W network remains passive, the closed loop gain condition becomes the most critical source of energy consumption.

\subsection{Optoelectronic Nonlinearities}
\label{sec:optoelectronic}
Researchers have used a variety of implementions to realize nonlinear functions in the optical domain. All-optical approaches have utilized nonlinear effects in both fibers~\cite{Trillo:88,Nelson:1991aa,Asobe:1997aa} and on-chip resonators~\cite{Tait:12,Notomi:07,Nozaki:2010aa,Van:2002aa}. Other approaches include carrier nonlinearities in SOAs~\cite{Sokoloff:1993aa,Stubkjaer:2000aa,Vandoorne:2011aa}, plasmonics~\cite{Ren:2011aa,Kauranen:2012aa} intracavity semiconductor saturable absorbers~\cite{takahashi1996ultrafast,Selmi:2014} and graphene~\cite{Shastri:2016aa}. However, these approaches consistently exhibit several common limitations: they either require exotic fabrication processes to create, or large optical threshold powers to activate. This can greatly increase the energy consumption to a level that negates the advantages of using optics at all.

Laser neurons use a detector-transducer configuration, a common optoelectronic nonlinear device template explored in the literature~\cite{Romeira:2016aa,Krauskopf2003,Nahmias:2016,Tait:2017aa,Nozaki:18}. O/E/O models (i.e., involving optical to electrical conversion and vise versa) can exploit nonlinearities in detectors and modulators or lasers, but are constrained by electrical parasitics and the costs of O/E and E/O conversion.
However, electro-optic conversion costs are shrinking: high performance detectors~\cite{Chen:09,Michel:2010aa}, modulators~\cite{Dong:09,Timurdogan:2014aa} and lasers~\cite{Wang:2015aa,Liu:15} are continuing to emerge in developing PIC platforms. In addition, placing detectors and transducers in close proximity can greatly minimize undesirable parasitics, including dispersion, microwave reflections, and timing delays~\cite{lee2003design}.

Another exciting prospect is the close integration between developing microelectronic and photonic platforms, which could combine a high O/E/O conversion efficiency with powerful nonlinear electronic operations. For example, operational amplifiers can be placed after detectors, compensating for loss and leading to greater system-level energy efficiency. Such hybrid units could potentially perform generic nonlinear tasks such as wavelength conversion very efficiently~\cite{Nozaki:18} and may provide ample machinery for nonlinear neural network processing in future systems.

\subsection{Spiking}

\label{sec:spiking}

Spiking is a communication encoding strategy that is equivalent to analog \emph{pulse position modulation} in photonics~\cite{Smith:1998aa,Sushchik:2000aa}: information is primarily encoded in the timing between a series of pulses, or \emph{spikes}. Spike amplitudes are binarized (i.e., either 0 or 1), but the timing of each pulse can take on any analog value. For example, an analog vector $[v_0, v_1, \dots ]$ can be encoded by associating each value $v_i$ with the time $T_0, T_1 \dots$ between each pulse, wherein the amount of information encoded is limited by the temporal resolution or timing jitter of the communication system.

Spike encoding has many advantages over continuous wave signals. For one, it is less susceptible to amplitude noise,
which can be useful in physical systems with high stochasticity (i.e., biology). Secondly, it benefits from sparse coding, which can lead to significant power advantages for photonic signals. As an illustrative example, for temporal resolution $\Delta t$, and delay $T_i$ between pulse $i$ and $i-1$, if $T_i > \Delta t$, a single pulse can carry more than 1 bit of information: as much as $N_b =
\log_2 (T_i/\Delta t)$. This reduces the \SI{}{J/bit} cost by a factor $N_b$ in a communication channel (see for example Ref.~\cite{Shiu:1999aa}), which can also improve the implementation of operations, such as MACs, in neuromorphic photonic systems.

Unfortunately, although spiking neural networks in hardware have continued to show significant power efficiency gains over other neural network processors~\cite{Merolla668,Furber:2016}, they are currently challenging to program and train. Spike-based learning algorithms---such as synaptic time dependent plasticity (STDP)---have difficulty propagating gradients back many layers, a prerequisite condition for deep learning. Nonetheless, improvements in this arena remains an active research topic~\cite{Pfeiffer:2018aa}, with many results on spiking training appearing recently~\cite{Samadi:2017aa,Kheradpisheh:2018,Tavanaei:2018}. As more sophisticated techniques are developed for the control of spiking neural networks, machine learning approaches may one day reap the robustness and energy efficiency that such encoding can deliver.


\section{Methods}
\subsection{Experimental Signal Generation}
\label{sec:experimental}

\begin{figure*}
\centering
\includegraphics[width=\linewidth]{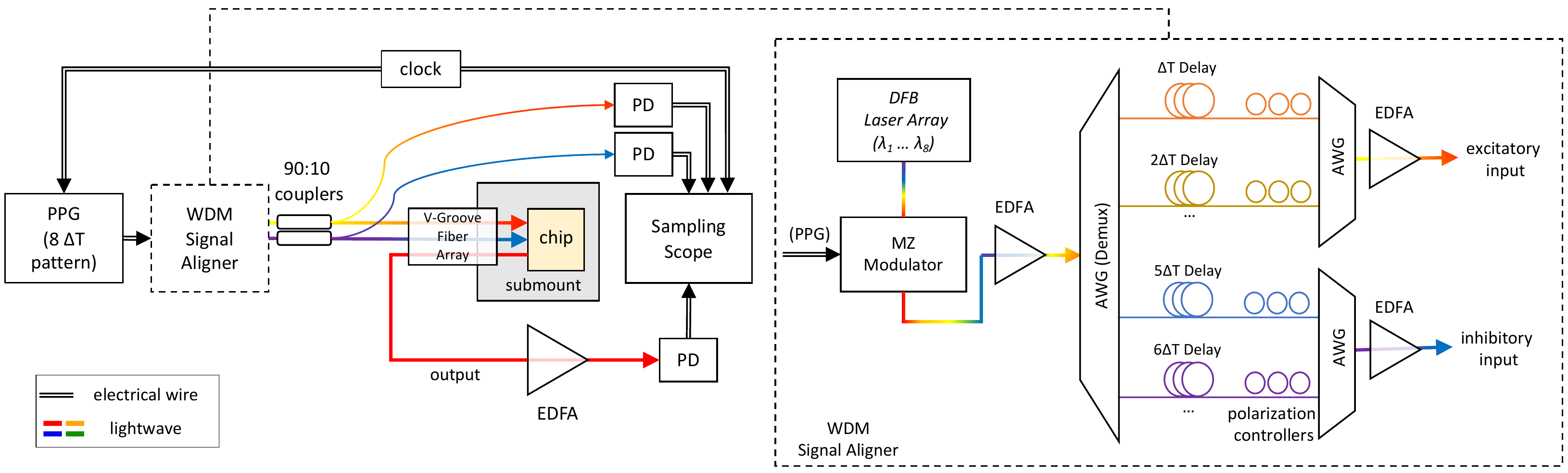}
\caption{Experimental set-up. All $8$ wavelength signals across a $\Delta T$ time interval are encoded in consecutive bit patterns across $8 \Delta T$ using a pulse pattern generator (PPG). These channels are later aligned using delay lines embedded within nested arrayed waveguide grating (AWG) multiplexer and demultiplexers. EDFAs are used to compensate for passive losses at the chip coupling interface and through the AWGs.}
\label{fig:input-gen}
\end{figure*}

Excitatory and inhibitory pulses were programmed using a custom input generation system, allowing $N$ intensity-modulated bit patterns along different wavelengths $\lambda_1, \lambda_2, \dots \lambda_{N}$ in some desired test window $t \in [0,\Delta T]$.
As illustrated in Fig.~\ref{fig:input-gen}, it consisted of a WDM source (an external array of DFB lasers), a single high speed Mach Zehnder (MZ) modulator connected to a pulse pattern generator (PPG) driven by a clock source, and a series of long delay lines nested between arrayed waveguide gratings (AWGs). The outputs were measured by a sampling scope connected to the same clock source. The system has many similarities to the generation mechanisms explored in Ref.~\cite{Tait:15,FerreiradeLima:2015}. We describe its mechanism of operation below.

Suppose we have a series of desired bit patterns denoted by $P_1[t], P_2[t], P_3[t] \dots P_N[t] $ along wavelength channels $k = 1 \dots N$ that include bit values $(0,1)$ programmed in the time interval $t \in [0,\Delta T]$. First, we set the PPG bit pattern as each desired pattern consecutively in time, i.e.,
\begin{align*}
\mathrm{PPG}[t] = P_k[t-(k-1)\Delta T] \text{ for } t \in [(k-1) \Delta T, k\Delta T]
\end{align*}
for $k = 1 \dots N$. The total length of the PPG bit pattern is therefore $[0,N \Delta T]$. This pattern is modulated onto all wavelengths $\lambda_k$ using a wideband modulator simultaneously.  Next, we send the signal through a Demux AWG and apply consecutive physical time delays $k \Delta T$ to each wavelength channel $k$ to cancel out the programmed time delays in the PPG. As a result, we generate our desired pattern in the time interval $[0,T]$:
\begin{align}
P_{in}[t] = \sum_{k=1}^{N} P_k[t] \text{ at } \lambda_k
\end{align}

In this experiment, we represented $N = 8$ wavelength channels using long fiber delays at intervals of $D_1 \sim \SI{90}{ns}$, $D_2 \sim \SI{180}{ns}$, etc.
Small variations in the physical delay for each fiber (around $\sigma \sim \SI{3}{ns}$) that resulted from splicing errors were compensated for digitally in the PPG delay time, i.e.,
\begin{align*}
\mathrm{PPG}[t] &= P_k[t-D_k] \text{ for } t \in [D_k, D_k + \Delta T]
\end{align*}
for physically measured delays $D_1 \dots D_N$, where we set the time window to $\Delta T = \min{(D_k-D_{k-1})}$ for all channels $k$ to avoid overlapping ($\Delta T \sim \SI{88}{ns}$).
The delayed output signals were multiplexed onto two output fibers using another set of AWGs, shown as the excitatory and inhibitory inputs in Fig.~\ref{fig:input-gen}. Three erbium doped fiber amplifiers (EDFAs) were placed in various parts of the signal pathway to compensate for losses: one after the MZ modulator and before the Demux AWG, and one for each excitatory and inhibitory input channel after each Mux AWG. The resulting fiber channels were input into a V-groove fiber array, which interfaced with both the inputs and output of each laser neuron via spot size converters (SSCs) on the edge of the chip. The output also received amplification via an EDFA to compensate for chip coupling losses.

Before inputs were coupled into the chip, 90:10 couplers were placed after each EDFA, wherein the smaller signals act both as power monitors and as outputs for measurements. Excitatory inputs shown in Fig.~\ref{fig:traces} were set at $\lambda_1 = 1540.56, \lambda_2 = 1543.72, \lambda_3 = 1546.92, \lambda_4 = 1550.12$, while inhibitory inputs are set at $\lambda_5 = 1542.14, \lambda_6 = 1545.32, \lambda_7 = 1548.52, \lambda_8 = 1551.72$. The output of each spiking laser neuron was measured using both sampling scope (for time-dependent traces in Figs~\ref{fig:traces}~and~\ref{fig:cascade}) and a spectrum analyzer (for spectral measurements in Fig~\ref{fig:spectrum}).

To generate Poisson inputs for the experiments conducted in Sec.~\ref{sec:cascadability}, we used only one wavelength input ($\lambda_4 = 1550.12$) fed into the excitatory port of a type {\bf A} neuron. The Poisson model was set with $\mu_p = \SI{1}{GHz}$ and a clock rate of \SI{5}{GHz} (each bit has $\tau_p = \SI{0.2}{ns}$) over a \SI{50}{ns} time window, which was generated a priori before being programmed into the PPG. The result was modulated onto a signal carrier wave at $\lambda_4 = 1550.12$. For both experiments, input and output powers traveling in and out of both the laser and BPD pair were calculated using a power calibration procedure based on the measured losses between the SSCs and fiber V-groove arrays.

\subsection{Data Analysis}

The powers of time-dependent traces were calibrated using a set of reference traces normalized via continuous wave power measurements. Up to three reference traces existed for each experiment: total excitatory input power, total inhibitory input power, and total laser output power, measured across the entire time window (i.e., $[0,N \Delta T]$). Once reference traces become normalized to average power measurements, all remaining data sets in the window of interest $[0, \Delta T]$ were calibrated to this reference set. Time-dependent traces for each wavelength $\lambda_i$ and the laser output were measured independently before calibration. This resulted in the final data plot seen in Fig. 5 and Fig. 6 in the main article (for which the latter plot used only one input wavelength channel).

\section{Data Availability}
The datasets generated during and/or analysed for this experiment are available from the corresponding author on reasonable request.

\bibliography{laser-spiking.bib}

\begin{thebibliography}{86}%
\makeatletter
\providecommand \@ifxundefined [1]{%
 \@ifx{#1\undefined}
}%
\providecommand \@ifnum [1]{%
 \ifnum #1\expandafter \@firstoftwo
 \else \expandafter \@secondoftwo
 \fi
}%
\providecommand \@ifx [1]{%
 \ifx #1\expandafter \@firstoftwo
 \else \expandafter \@secondoftwo
 \fi
}%
\providecommand \natexlab [1]{#1}%
\providecommand \enquote  [1]{``#1''}%
\providecommand \bibnamefont  [1]{#1}%
\providecommand \bibfnamefont [1]{#1}%
\providecommand \citenamefont [1]{#1}%
\providecommand \href@noop [0]{\@secondoftwo}%
\providecommand \href [0]{\begingroup \@sanitize@url \@href}%
\providecommand \@href[1]{\@@startlink{#1}\@@href}%
\providecommand \@@href[1]{\endgroup#1\@@endlink}%
\providecommand \@sanitize@url [0]{\catcode `\\12\catcode `\$12\catcode
  `\&12\catcode `\#12\catcode `\^12\catcode `\_12\catcode `\%12\relax}%
\providecommand \@@startlink[1]{}%
\providecommand \@@endlink[0]{}%
\providecommand \url  [0]{\begingroup\@sanitize@url \@url }%
\providecommand \@url [1]{\endgroup\@href {#1}{\urlprefix }}%
\providecommand \urlprefix  [0]{URL }%
\providecommand \Eprint [0]{\href }%
\providecommand \doibase [0]{http://dx.doi.org/}%
\providecommand \selectlanguage [0]{\@gobble}%
\providecommand \bibinfo  [0]{\@secondoftwo}%
\providecommand \bibfield  [0]{\@secondoftwo}%
\providecommand \translation [1]{[#1]}%
\providecommand \BibitemOpen [0]{}%
\providecommand \bibitemStop [0]{}%
\providecommand \bibitemNoStop [0]{.\EOS\space}%
\providecommand \EOS [0]{\spacefactor3000\relax}%
\providecommand \BibitemShut  [1]{\csname bibitem#1\endcsname}%
\let\auto@bib@innerbib\@empty
\bibitem [{\citenamefont {Amodei}\ and\ \citenamefont
  {Hernandez}(2018)}]{openai-compute}%
  \BibitemOpen
  \bibfield  {author} {\bibinfo {author} {\bibfnamefont {D.}~\bibnamefont
  {Amodei}}\ and\ \bibinfo {author} {\bibfnamefont {D.}~\bibnamefont
  {Hernandez}},\ }\href {https://blog.openai.com/ai-and-compute/} {\enquote
  {\bibinfo {title} {Ai and compute},}\ } (\bibinfo {year} {2018})\BibitemShut
  {NoStop}%
\bibitem [{\citenamefont {Esmaeilzadeh}\ \emph {et~al.}(2012)\citenamefont
  {Esmaeilzadeh}, \citenamefont {Blem}, \citenamefont {{St. Amant}},
  \citenamefont {Sankaralingam},\ and\ \citenamefont
  {Burger}}]{Esmaeilzadeh2012}%
  \BibitemOpen
  \bibfield  {author} {\bibinfo {author} {\bibfnamefont {H.}~\bibnamefont
  {Esmaeilzadeh}}, \bibinfo {author} {\bibfnamefont {E.}~\bibnamefont {Blem}},
  \bibinfo {author} {\bibfnamefont {R.}~\bibnamefont {{St. Amant}}}, \bibinfo
  {author} {\bibfnamefont {K.}~\bibnamefont {Sankaralingam}}, \ and\ \bibinfo
  {author} {\bibfnamefont {D.}~\bibnamefont {Burger}},\ }\bibfield  {title}
  {\enquote {\bibinfo {title} {{Dark Silicon and the End of Multicore
  Scaling}},}\ }\href {\doibase 10.1109/MM.2012.17} {\bibfield  {journal}
  {\bibinfo  {journal} {IEEE Micro}\ }\textbf {\bibinfo {volume} {32}},\
  \bibinfo {pages} {122--134} (\bibinfo {year} {2012})}\BibitemShut {NoStop}%
\bibitem [{ird(2017)}]{irds2017}%
  \BibitemOpen
  \href {https://irds.ieee.org/roadmap-2017} {\enquote {\bibinfo {title}
  {International roadmap for devices and systems 2017 edition},}\ }\bibinfo
  {type} {Tech. Rep.}\ (\bibinfo  {institution} {Institute of Electrical and
  Electronics Engineers},\ \bibinfo {year} {2017})\BibitemShut {NoStop}%
\bibitem [{\citenamefont {et~al.}(2017)}]{Jouppi:2017aa}%
  \BibitemOpen
  \bibfield  {author} {\bibinfo {author} {\bibfnamefont {N.~P.~J.}\
  \bibnamefont {et~al.}},\ }\bibfield  {title} {\enquote {\bibinfo {title}
  {In-datacenter performance analysis of a tensor processing unit},}\ }in\
  \href {\doibase 10.1145/3079856.3080246} {\emph {\bibinfo {booktitle} {2017
  ACM/IEEE 44th Annual International Symposium on Computer Architecture
  (ISCA)}}}\ (\bibinfo {year} {2017})\ pp.\ \bibinfo {pages}
  {1--12}\BibitemShut {NoStop}%
\bibitem [{\citenamefont {Yu}(2018)}]{Yu:2018aa}%
  \BibitemOpen
  \bibfield  {author} {\bibinfo {author} {\bibfnamefont {S.}~\bibnamefont
  {Yu}},\ }\bibfield  {title} {\enquote {\bibinfo {title} {Neuro-inspired
  computing with emerging nonvolatile memorys},}\ }\bibfield  {booktitle}
  {\emph {\bibinfo {booktitle} {Proceedings of the IEEE}},\ }\href {\doibase
  10.1109/JPROC.2018.2790840} {\bibfield  {journal} {\bibinfo  {journal}
  {Proceedings of the IEEE}\ }\textbf {\bibinfo {volume} {106}},\ \bibinfo
  {pages} {260--285} (\bibinfo {year} {2018})}\BibitemShut {NoStop}%
\bibitem [{\citenamefont {Ielmini}\ and\ \citenamefont
  {Wong}(2018)}]{Ielmini:2018aa}%
  \BibitemOpen
  \bibfield  {author} {\bibinfo {author} {\bibfnamefont {D.}~\bibnamefont
  {Ielmini}}\ and\ \bibinfo {author} {\bibfnamefont {H.~S.~P.}\ \bibnamefont
  {Wong}},\ }\bibfield  {title} {\enquote {\bibinfo {title} {In-memory
  computing with resistive switching devices},}\ }\href {\doibase
  10.1038/s41928-018-0092-2} {\bibfield  {journal} {\bibinfo  {journal} {Nature
  Electronics}\ }\textbf {\bibinfo {volume} {1}},\ \bibinfo {pages} {333--343}
  (\bibinfo {year} {2018})}\BibitemShut {NoStop}%
\bibitem [{\citenamefont {Burr}\ \emph {et~al.}(2017)\citenamefont {Burr},
  \citenamefont {Shelby}, \citenamefont {Sebastian}, \citenamefont {Kim},
  \citenamefont {Kim}, \citenamefont {Sidler}, \citenamefont {Virwani},
  \citenamefont {Ishii}, \citenamefont {Narayanan}, \citenamefont {Fumarola},
  \citenamefont {Sanches}, \citenamefont {Boybat}, \citenamefont {Gallo},
  \citenamefont {Moon}, \citenamefont {Woo}, \citenamefont {Hwang},\ and\
  \citenamefont {Leblebici}}]{Burr:2017}%
  \BibitemOpen
  \bibfield  {author} {\bibinfo {author} {\bibfnamefont {G.~W.}\ \bibnamefont
  {Burr}}, \bibinfo {author} {\bibfnamefont {R.~M.}\ \bibnamefont {Shelby}},
  \bibinfo {author} {\bibfnamefont {A.}~\bibnamefont {Sebastian}}, \bibinfo
  {author} {\bibfnamefont {S.}~\bibnamefont {Kim}}, \bibinfo {author}
  {\bibfnamefont {S.}~\bibnamefont {Kim}}, \bibinfo {author} {\bibfnamefont
  {S.}~\bibnamefont {Sidler}}, \bibinfo {author} {\bibfnamefont
  {K.}~\bibnamefont {Virwani}}, \bibinfo {author} {\bibfnamefont
  {M.}~\bibnamefont {Ishii}}, \bibinfo {author} {\bibfnamefont
  {P.}~\bibnamefont {Narayanan}}, \bibinfo {author} {\bibfnamefont
  {A.}~\bibnamefont {Fumarola}}, \bibinfo {author} {\bibfnamefont {L.~L.}\
  \bibnamefont {Sanches}}, \bibinfo {author} {\bibfnamefont {I.}~\bibnamefont
  {Boybat}}, \bibinfo {author} {\bibfnamefont {M.~L.}\ \bibnamefont {Gallo}},
  \bibinfo {author} {\bibfnamefont {K.}~\bibnamefont {Moon}}, \bibinfo {author}
  {\bibfnamefont {J.}~\bibnamefont {Woo}}, \bibinfo {author} {\bibfnamefont
  {H.}~\bibnamefont {Hwang}}, \ and\ \bibinfo {author} {\bibfnamefont
  {Y.}~\bibnamefont {Leblebici}},\ }\bibfield  {title} {\enquote {\bibinfo
  {title} {Neuromorphic computing using non-volatile memory},}\ }\href
  {\doibase 10.1080/23746149.2016.1259585} {\bibfield  {journal} {\bibinfo
  {journal} {Advances in Physics: X}\ }\textbf {\bibinfo {volume} {2}},\
  \bibinfo {pages} {89--124} (\bibinfo {year} {2017})},\ \Eprint
  {http://arxiv.org/abs/https://doi.org/10.1080/23746149.2016.1259585}
  {https://doi.org/10.1080/23746149.2016.1259585} \BibitemShut {NoStop}%
\bibitem [{\citenamefont {Ambrogio}\ \emph {et~al.}(2018)\citenamefont
  {Ambrogio}, \citenamefont {Narayanan}, \citenamefont {Tsai}, \citenamefont
  {Shelby}, \citenamefont {Boybat}, \citenamefont {di~Nolfo}, \citenamefont
  {Sidler}, \citenamefont {Giordano}, \citenamefont {Bodini}, \citenamefont
  {Farinha}, \citenamefont {Killeen}, \citenamefont {Cheng}, \citenamefont
  {Jaoudi},\ and\ \citenamefont {Burr}}]{Ambrogio:2018aa}%
  \BibitemOpen
  \bibfield  {author} {\bibinfo {author} {\bibfnamefont {S.}~\bibnamefont
  {Ambrogio}}, \bibinfo {author} {\bibfnamefont {P.}~\bibnamefont {Narayanan}},
  \bibinfo {author} {\bibfnamefont {H.}~\bibnamefont {Tsai}}, \bibinfo {author}
  {\bibfnamefont {R.~M.}\ \bibnamefont {Shelby}}, \bibinfo {author}
  {\bibfnamefont {I.}~\bibnamefont {Boybat}}, \bibinfo {author} {\bibfnamefont
  {C.}~\bibnamefont {di~Nolfo}}, \bibinfo {author} {\bibfnamefont
  {S.}~\bibnamefont {Sidler}}, \bibinfo {author} {\bibfnamefont
  {M.}~\bibnamefont {Giordano}}, \bibinfo {author} {\bibfnamefont
  {M.}~\bibnamefont {Bodini}}, \bibinfo {author} {\bibfnamefont {N.~C.~P.}\
  \bibnamefont {Farinha}}, \bibinfo {author} {\bibfnamefont {B.}~\bibnamefont
  {Killeen}}, \bibinfo {author} {\bibfnamefont {C.}~\bibnamefont {Cheng}},
  \bibinfo {author} {\bibfnamefont {Y.}~\bibnamefont {Jaoudi}}, \ and\ \bibinfo
  {author} {\bibfnamefont {G.~W.}\ \bibnamefont {Burr}},\ }\bibfield  {title}
  {\enquote {\bibinfo {title} {Equivalent-accuracy accelerated neural-network
  training using analogue memory},}\ }\href {\doibase
  10.1038/s41586-018-0180-5} {\bibfield  {journal} {\bibinfo  {journal}
  {Nature}\ }\textbf {\bibinfo {volume} {558}},\ \bibinfo {pages} {60--67}
  (\bibinfo {year} {2018})}\BibitemShut {NoStop}%
\bibitem [{\citenamefont {Jo}\ \emph {et~al.}(2010)\citenamefont {Jo},
  \citenamefont {Chang}, \citenamefont {Ebong}, \citenamefont {Bhadviya},
  \citenamefont {Mazumder},\ and\ \citenamefont {Lu}}]{Jo:2010}%
  \BibitemOpen
  \bibfield  {author} {\bibinfo {author} {\bibfnamefont {S.~H.}\ \bibnamefont
  {Jo}}, \bibinfo {author} {\bibfnamefont {T.}~\bibnamefont {Chang}}, \bibinfo
  {author} {\bibfnamefont {I.}~\bibnamefont {Ebong}}, \bibinfo {author}
  {\bibfnamefont {B.~B.}\ \bibnamefont {Bhadviya}}, \bibinfo {author}
  {\bibfnamefont {P.}~\bibnamefont {Mazumder}}, \ and\ \bibinfo {author}
  {\bibfnamefont {W.}~\bibnamefont {Lu}},\ }\bibfield  {title} {\enquote
  {\bibinfo {title} {Nanoscale memristor device as synapse in neuromorphic
  systems},}\ }\href@noop {} {\bibfield  {journal} {\bibinfo  {journal} {Nano
  Letters}\ }\textbf {\bibinfo {volume} {10}},\ \bibinfo {pages} {1297--1301}
  (\bibinfo {year} {2010})}\BibitemShut {NoStop}%
\bibitem [{\citenamefont {Prezioso}\ \emph {et~al.}(2015)\citenamefont
  {Prezioso}, \citenamefont {Merrikh-Bayat}, \citenamefont {Hoskins},
  \citenamefont {Adam}, \citenamefont {Likharev},\ and\ \citenamefont
  {Strukov}}]{Prezioso:2015aa}%
  \BibitemOpen
  \bibfield  {author} {\bibinfo {author} {\bibfnamefont {M.}~\bibnamefont
  {Prezioso}}, \bibinfo {author} {\bibfnamefont {F.}~\bibnamefont
  {Merrikh-Bayat}}, \bibinfo {author} {\bibfnamefont {B.~D.}\ \bibnamefont
  {Hoskins}}, \bibinfo {author} {\bibfnamefont {G.~C.}\ \bibnamefont {Adam}},
  \bibinfo {author} {\bibfnamefont {K.~K.}\ \bibnamefont {Likharev}}, \ and\
  \bibinfo {author} {\bibfnamefont {D.~B.}\ \bibnamefont {Strukov}},\
  }\bibfield  {title} {\enquote {\bibinfo {title} {Training and operation of an
  integrated neuromorphic network based on metal-oxide memristors},}\ }\href
  {https://doi.org/10.1038/nature14441} {\bibfield  {journal} {\bibinfo
  {journal} {Nature}\ }\textbf {\bibinfo {volume} {521}},\ \bibinfo {pages} {61
  EP --} (\bibinfo {year} {2015})}\BibitemShut {NoStop}%
\bibitem [{\citenamefont {Bayat}\ \emph {et~al.}(2018)\citenamefont {Bayat},
  \citenamefont {Prezioso}, \citenamefont {Chakrabarti}, \citenamefont {Nili},
  \citenamefont {Kataeva},\ and\ \citenamefont {Strukov}}]{Bayat:2018aa}%
  \BibitemOpen
  \bibfield  {author} {\bibinfo {author} {\bibfnamefont {F.~M.}\ \bibnamefont
  {Bayat}}, \bibinfo {author} {\bibfnamefont {M.}~\bibnamefont {Prezioso}},
  \bibinfo {author} {\bibfnamefont {B.}~\bibnamefont {Chakrabarti}}, \bibinfo
  {author} {\bibfnamefont {H.}~\bibnamefont {Nili}}, \bibinfo {author}
  {\bibfnamefont {I.}~\bibnamefont {Kataeva}}, \ and\ \bibinfo {author}
  {\bibfnamefont {D.}~\bibnamefont {Strukov}},\ }\bibfield  {title} {\enquote
  {\bibinfo {title} {Implementation of multilayer perceptron network with
  highly uniform passive memristive crossbar circuits},}\ }\href {\doibase
  10.1038/s41467-018-04482-4} {\bibfield  {journal} {\bibinfo  {journal}
  {Nature Communications}\ }\textbf {\bibinfo {volume} {9}},\ \bibinfo {pages}
  {2331} (\bibinfo {year} {2018})}\BibitemShut {NoStop}%
\bibitem [{\citenamefont {Miller}(2009)}]{miller2009device}%
  \BibitemOpen
  \bibfield  {author} {\bibinfo {author} {\bibfnamefont {D.~A.~B.}\
  \bibnamefont {Miller}},\ }\bibfield  {title} {\enquote {\bibinfo {title}
  {Device requirements for optical interconnects to silicon chips},}\
  }\href@noop {} {\bibfield  {journal} {\bibinfo  {journal} {Proceedings of the
  IEEE}\ }\textbf {\bibinfo {volume} {97}},\ \bibinfo {pages} {1166--1185}
  (\bibinfo {year} {2009})}\BibitemShut {NoStop}%
\bibitem [{\citenamefont {Psaltis}, \citenamefont {Brady},\ and\ \citenamefont
  {Wagner}(1988)}]{Psaltis:88}%
  \BibitemOpen
  \bibfield  {author} {\bibinfo {author} {\bibfnamefont {D.}~\bibnamefont
  {Psaltis}}, \bibinfo {author} {\bibfnamefont {D.}~\bibnamefont {Brady}}, \
  and\ \bibinfo {author} {\bibfnamefont {K.}~\bibnamefont {Wagner}},\
  }\bibfield  {title} {\enquote {\bibinfo {title} {Adaptive optical networks
  using photorefractive crystals},}\ }\href {\doibase 10.1364/AO.27.001752}
  {\bibfield  {journal} {\bibinfo  {journal} {Appl. Opt.}\ }\textbf {\bibinfo
  {volume} {27}},\ \bibinfo {pages} {1752--1759} (\bibinfo {year}
  {1988})}\BibitemShut {NoStop}%
\bibitem [{\citenamefont {Georgas}\ \emph {et~al.}(2011)\citenamefont
  {Georgas}, \citenamefont {Leu}, \citenamefont {Moss}, \citenamefont {Sun},\
  and\ \citenamefont {Stojanovi{\'c}}}]{Georgas:2011aa}%
  \BibitemOpen
  \bibfield  {author} {\bibinfo {author} {\bibfnamefont {M.}~\bibnamefont
  {Georgas}}, \bibinfo {author} {\bibfnamefont {J.}~\bibnamefont {Leu}},
  \bibinfo {author} {\bibfnamefont {B.}~\bibnamefont {Moss}}, \bibinfo {author}
  {\bibfnamefont {C.}~\bibnamefont {Sun}}, \ and\ \bibinfo {author}
  {\bibfnamefont {V.}~\bibnamefont {Stojanovi{\'c}}},\ }\bibfield  {title}
  {\enquote {\bibinfo {title} {Addressing link-level design tradeoffs for
  integrated photonic interconnects},}\ }in\ \href {\doibase
  10.1109/CICC.2011.6055363} {\emph {\bibinfo {booktitle} {2011 IEEE Custom
  Integrated Circuits Conference (CICC)}}}\ (\bibinfo {year} {2011})\ pp.\
  \bibinfo {pages} {1--8}\BibitemShut {NoStop}%
\bibitem [{\citenamefont {Miller}(2017)}]{Miller:17}%
  \BibitemOpen
  \bibfield  {author} {\bibinfo {author} {\bibfnamefont {D.~A.~B.}\
  \bibnamefont {Miller}},\ }\bibfield  {title} {\enquote {\bibinfo {title}
  {Attojoule optoelectronics for low-energy information processing and
  communications},}\ }\href {http://jlt.osa.org/abstract.cfm?URI=jlt-35-3-346}
  {\bibfield  {journal} {\bibinfo  {journal} {J. Lightwave Technol.}\ }\textbf
  {\bibinfo {volume} {35}},\ \bibinfo {pages} {346--396} (\bibinfo {year}
  {2017})}\BibitemShut {NoStop}%
\bibitem [{\citenamefont {Agarwal}\ \emph {et~al.}(2016)\citenamefont
  {Agarwal}, \citenamefont {Quach}, \citenamefont {Parekh}, \citenamefont
  {Hsia}, \citenamefont {DeBenedictis}, \citenamefont {James}, \citenamefont
  {Marinella},\ and\ \citenamefont {Aimone}}]{10.3389/fnins.2015.00484}%
  \BibitemOpen
  \bibfield  {author} {\bibinfo {author} {\bibfnamefont {S.}~\bibnamefont
  {Agarwal}}, \bibinfo {author} {\bibfnamefont {T.-T.}\ \bibnamefont {Quach}},
  \bibinfo {author} {\bibfnamefont {O.}~\bibnamefont {Parekh}}, \bibinfo
  {author} {\bibfnamefont {A.~H.}\ \bibnamefont {Hsia}}, \bibinfo {author}
  {\bibfnamefont {E.~P.}\ \bibnamefont {DeBenedictis}}, \bibinfo {author}
  {\bibfnamefont {C.~D.}\ \bibnamefont {James}}, \bibinfo {author}
  {\bibfnamefont {M.~J.}\ \bibnamefont {Marinella}}, \ and\ \bibinfo {author}
  {\bibfnamefont {J.~B.}\ \bibnamefont {Aimone}},\ }\bibfield  {title}
  {\enquote {\bibinfo {title} {Energy scaling advantages of resistive memory
  crossbar based computation and its application to sparse coding},}\ }\href
  {\doibase 10.3389/fnins.2015.00484} {\bibfield  {journal} {\bibinfo
  {journal} {Frontiers in Neuroscience}\ }\textbf {\bibinfo {volume} {9}},\
  \bibinfo {pages} {484} (\bibinfo {year} {2016})}\BibitemShut {NoStop}%
\bibitem [{\citenamefont {Shen}\ \emph {et~al.}(2017)\citenamefont {Shen},
  \citenamefont {Harris}, \citenamefont {Skirlo}, \citenamefont {Prabhu},
  \citenamefont {Baehr-Jones}, \citenamefont {Hochberg}, \citenamefont {Sun},
  \citenamefont {Zhao}, \citenamefont {Larochelle}, \citenamefont {Englund},\
  and\ \citenamefont {Solja{\v c}i{\'c}}}]{Shen:2017aa}%
  \BibitemOpen
  \bibfield  {author} {\bibinfo {author} {\bibfnamefont {Y.}~\bibnamefont
  {Shen}}, \bibinfo {author} {\bibfnamefont {N.~C.}\ \bibnamefont {Harris}},
  \bibinfo {author} {\bibfnamefont {S.}~\bibnamefont {Skirlo}}, \bibinfo
  {author} {\bibfnamefont {M.}~\bibnamefont {Prabhu}}, \bibinfo {author}
  {\bibfnamefont {T.}~\bibnamefont {Baehr-Jones}}, \bibinfo {author}
  {\bibfnamefont {M.}~\bibnamefont {Hochberg}}, \bibinfo {author}
  {\bibfnamefont {X.}~\bibnamefont {Sun}}, \bibinfo {author} {\bibfnamefont
  {S.}~\bibnamefont {Zhao}}, \bibinfo {author} {\bibfnamefont {H.}~\bibnamefont
  {Larochelle}}, \bibinfo {author} {\bibfnamefont {D.}~\bibnamefont {Englund}},
  \ and\ \bibinfo {author} {\bibfnamefont {M.}~\bibnamefont {Solja{\v
  c}i{\'c}}},\ }\bibfield  {title} {\enquote {\bibinfo {title} {Deep learning
  with coherent nanophotonic circuits},}\ }\href
  {http://dx.doi.org/10.1038/nphoton.2017.93} {\bibfield  {journal} {\bibinfo
  {journal} {Nature Photonics}\ }\textbf {\bibinfo {volume} {11}},\ \bibinfo
  {pages} {441 EP --} (\bibinfo {year} {2017})}\BibitemShut {NoStop}%
\bibitem [{\citenamefont {Prucnal}\ and\ \citenamefont
  {Shastri}(2017)}]{PrucnalBook}%
  \BibitemOpen
  \bibfield  {author} {\bibinfo {author} {\bibfnamefont {P.~R.}\ \bibnamefont
  {Prucnal}}\ and\ \bibinfo {author} {\bibfnamefont {B.~J.}\ \bibnamefont
  {Shastri}},\ }\href@noop {} {\emph {\bibinfo {title} {{Neuromorphic
  Photonics}}}}\ (\bibinfo  {publisher} {CRC Press, Taylor \& Francis Group},\
  \bibinfo {address} {Boca Raton, FL, USA},\ \bibinfo {year}
  {2017})\BibitemShut {NoStop}%
\bibitem [{\citenamefont {Walden}(1999)}]{Walden:1999aa}%
  \BibitemOpen
  \bibfield  {author} {\bibinfo {author} {\bibfnamefont {R.~H.}\ \bibnamefont
  {Walden}},\ }\bibfield  {title} {\enquote {\bibinfo {title} {Performance
  trends for analog to digital converters},}\ }\bibfield  {booktitle} {\emph
  {\bibinfo {booktitle} {IEEE Communications Magazine}},\ }\href {\doibase
  10.1109/35.747256} {\bibfield  {journal} {\bibinfo  {journal} {IEEE
  Communications Magazine}\ }\textbf {\bibinfo {volume} {37}},\ \bibinfo
  {pages} {96--101} (\bibinfo {year} {1999})}\BibitemShut {NoStop}%
\bibitem [{\citenamefont {Walden}(2008)}]{Walden:2008}%
  \BibitemOpen
  \bibfield  {author} {\bibinfo {author} {\bibfnamefont {R.~H.}\ \bibnamefont
  {Walden}},\ }\bibfield  {title} {\enquote {\bibinfo {title}
  {Analog-to-digital conversion in the early twenty-first century},}\
  }\bibfield  {booktitle} {\emph {\bibinfo {booktitle} {Wiley Encyclopedia of
  Computer Science and Engineering}},\ }\href {\doibase
  10.1002/9780470050118.ecse014} {\ ,\ \bibinfo {pages} {1--14} (\bibinfo
  {year} {2008})}\BibitemShut {NoStop}%
\bibitem [{\citenamefont {Tait}\ \emph {et~al.}(2013)\citenamefont {Tait},
  \citenamefont {Shastri}, \citenamefont {Fok}, \citenamefont {Nahmias},\ and\
  \citenamefont {Prucnal}}]{Tait:2013}%
  \BibitemOpen
  \bibfield  {author} {\bibinfo {author} {\bibfnamefont {A.~N.}\ \bibnamefont
  {Tait}}, \bibinfo {author} {\bibfnamefont {B.~J.}\ \bibnamefont {Shastri}},
  \bibinfo {author} {\bibfnamefont {M.~P.}\ \bibnamefont {Fok}}, \bibinfo
  {author} {\bibfnamefont {M.~A.}\ \bibnamefont {Nahmias}}, \ and\ \bibinfo
  {author} {\bibfnamefont {P.~R.}\ \bibnamefont {Prucnal}},\ }\bibfield
  {title} {\enquote {\bibinfo {title} {The dream: An integrated photonic
  thresholder},}\ }\href {\doibase 10.1109/JLT.2013.2246544} {\bibfield
  {journal} {\bibinfo  {journal} {Journal of Lightwave Technology}\ }\textbf
  {\bibinfo {volume} {31}},\ \bibinfo {pages} {1263--1272} (\bibinfo {year}
  {2013})}\BibitemShut {NoStop}%
\bibitem [{\citenamefont {Notomi}\ \emph {et~al.}(2007)\citenamefont {Notomi},
  \citenamefont {Tanabe}, \citenamefont {Shinya}, \citenamefont {Kuramochi},
  \citenamefont {Taniyama}, \citenamefont {Mitsugi},\ and\ \citenamefont
  {Morita}}]{Notomi:07}%
  \BibitemOpen
  \bibfield  {author} {\bibinfo {author} {\bibfnamefont {M.}~\bibnamefont
  {Notomi}}, \bibinfo {author} {\bibfnamefont {T.}~\bibnamefont {Tanabe}},
  \bibinfo {author} {\bibfnamefont {A.}~\bibnamefont {Shinya}}, \bibinfo
  {author} {\bibfnamefont {E.}~\bibnamefont {Kuramochi}}, \bibinfo {author}
  {\bibfnamefont {H.}~\bibnamefont {Taniyama}}, \bibinfo {author}
  {\bibfnamefont {S.}~\bibnamefont {Mitsugi}}, \ and\ \bibinfo {author}
  {\bibfnamefont {M.}~\bibnamefont {Morita}},\ }\bibfield  {title} {\enquote
  {\bibinfo {title} {Nonlinear and adiabatic control of high-q photonic crystal
  nanocavities},}\ }\href {\doibase 10.1364/OE.15.017458} {\bibfield  {journal}
  {\bibinfo  {journal} {Optics Express}\ }\textbf {\bibinfo {volume} {15}},\
  \bibinfo {pages} {17458--17481} (\bibinfo {year} {2007})}\BibitemShut
  {NoStop}%
\bibitem [{\citenamefont {Nozaki}\ \emph {et~al.}(2010)\citenamefont {Nozaki},
  \citenamefont {Tanabe}, \citenamefont {Shinya}, \citenamefont {Matsuo},
  \citenamefont {Sato}, \citenamefont {Taniyama},\ and\ \citenamefont
  {Notomi}}]{Nozaki:2010aa}%
  \BibitemOpen
  \bibfield  {author} {\bibinfo {author} {\bibfnamefont {K.}~\bibnamefont
  {Nozaki}}, \bibinfo {author} {\bibfnamefont {T.}~\bibnamefont {Tanabe}},
  \bibinfo {author} {\bibfnamefont {A.}~\bibnamefont {Shinya}}, \bibinfo
  {author} {\bibfnamefont {S.}~\bibnamefont {Matsuo}}, \bibinfo {author}
  {\bibfnamefont {T.}~\bibnamefont {Sato}}, \bibinfo {author} {\bibfnamefont
  {H.}~\bibnamefont {Taniyama}}, \ and\ \bibinfo {author} {\bibfnamefont
  {M.}~\bibnamefont {Notomi}},\ }\bibfield  {title} {\enquote {\bibinfo {title}
  {Sub-femtojoule all-optical switching using a photonic-crystal nanocavity},}\
  }\href {https://doi.org/10.1038/nphoton.2010.89} {\bibfield  {journal}
  {\bibinfo  {journal} {Nature Photonics}\ }\textbf {\bibinfo {volume} {4}},\
  \bibinfo {pages} {477 EP --} (\bibinfo {year} {2010})}\BibitemShut {NoStop}%
\bibitem [{\citenamefont {Van}\ \emph {et~al.}(2002)\citenamefont {Van},
  \citenamefont {Ibrahim}, \citenamefont {Ritter}, \citenamefont {Absil},
  \citenamefont {Johnson}, \citenamefont {Grover}, \citenamefont {Goldhar},\
  and\ \citenamefont {Ho}}]{Van:2002aa}%
  \BibitemOpen
  \bibfield  {author} {\bibinfo {author} {\bibfnamefont {V.}~\bibnamefont
  {Van}}, \bibinfo {author} {\bibfnamefont {T.~A.}\ \bibnamefont {Ibrahim}},
  \bibinfo {author} {\bibfnamefont {K.}~\bibnamefont {Ritter}}, \bibinfo
  {author} {\bibfnamefont {P.~P.}\ \bibnamefont {Absil}}, \bibinfo {author}
  {\bibfnamefont {F.~G.}\ \bibnamefont {Johnson}}, \bibinfo {author}
  {\bibfnamefont {R.}~\bibnamefont {Grover}}, \bibinfo {author} {\bibfnamefont
  {J.}~\bibnamefont {Goldhar}}, \ and\ \bibinfo {author} {\bibfnamefont {P.~.}\
  \bibnamefont {Ho}},\ }\bibfield  {title} {\enquote {\bibinfo {title}
  {All-optical nonlinear switching in gaas-algaas microring resonators},}\
  }\bibfield  {booktitle} {\emph {\bibinfo {booktitle} {IEEE Photonics
  Technology Letters}},\ }\href {\doibase 10.1109/68.974166} {\bibfield
  {journal} {\bibinfo  {journal} {IEEE Photonics Technology Letters}\ }\textbf
  {\bibinfo {volume} {14}},\ \bibinfo {pages} {74--76} (\bibinfo {year}
  {2002})}\BibitemShut {NoStop}%
\bibitem [{\citenamefont {Ren}\ \emph {et~al.}(2011)\citenamefont {Ren},
  \citenamefont {Jia}, \citenamefont {Ou}, \citenamefont {Plum}, \citenamefont
  {Zhang}, \citenamefont {MacDonald}, \citenamefont {Nikolaenko}, \citenamefont
  {Xu}, \citenamefont {Gu},\ and\ \citenamefont {Zheludev}}]{Ren:2011aa}%
  \BibitemOpen
  \bibfield  {author} {\bibinfo {author} {\bibfnamefont {M.}~\bibnamefont
  {Ren}}, \bibinfo {author} {\bibfnamefont {B.}~\bibnamefont {Jia}}, \bibinfo
  {author} {\bibfnamefont {J.-Y.}\ \bibnamefont {Ou}}, \bibinfo {author}
  {\bibfnamefont {E.}~\bibnamefont {Plum}}, \bibinfo {author} {\bibfnamefont
  {J.}~\bibnamefont {Zhang}}, \bibinfo {author} {\bibfnamefont {K.~F.}\
  \bibnamefont {MacDonald}}, \bibinfo {author} {\bibfnamefont {A.~E.}\
  \bibnamefont {Nikolaenko}}, \bibinfo {author} {\bibfnamefont
  {J.}~\bibnamefont {Xu}}, \bibinfo {author} {\bibfnamefont {M.}~\bibnamefont
  {Gu}}, \ and\ \bibinfo {author} {\bibfnamefont {N.~I.}\ \bibnamefont
  {Zheludev}},\ }\bibfield  {title} {\enquote {\bibinfo {title} {Nanostructured
  plasmonic medium for terahertz bandwidth all-optical switching},}\ }\bibfield
   {booktitle} {\emph {\bibinfo {booktitle} {Advanced Materials}},\ }\href
  {\doibase 10.1002/adma.201103162} {\bibfield  {journal} {\bibinfo  {journal}
  {Advanced Materials}\ }\textbf {\bibinfo {volume} {23}},\ \bibinfo {pages}
  {5540--5544} (\bibinfo {year} {2011})}\BibitemShut {NoStop}%
\bibitem [{\citenamefont {Kauranen}\ and\ \citenamefont
  {Zayats}(2012)}]{Kauranen:2012aa}%
  \BibitemOpen
  \bibfield  {author} {\bibinfo {author} {\bibfnamefont {M.}~\bibnamefont
  {Kauranen}}\ and\ \bibinfo {author} {\bibfnamefont {A.~V.}\ \bibnamefont
  {Zayats}},\ }\bibfield  {title} {\enquote {\bibinfo {title} {Nonlinear
  plasmonics},}\ }\href {https://doi.org/10.1038/nphoton.2012.244} {\bibfield
  {journal} {\bibinfo  {journal} {Nature Photonics}\ }\textbf {\bibinfo
  {volume} {6}},\ \bibinfo {pages} {737 EP --} (\bibinfo {year}
  {2012})}\BibitemShut {NoStop}%
\bibitem [{\citenamefont {Takahashi}, \citenamefont {Kawamura},\ and\
  \citenamefont {Iwamura}(1996)}]{takahashi1996ultrafast}%
  \BibitemOpen
  \bibfield  {author} {\bibinfo {author} {\bibfnamefont {R.}~\bibnamefont
  {Takahashi}}, \bibinfo {author} {\bibfnamefont {Y.}~\bibnamefont {Kawamura}},
  \ and\ \bibinfo {author} {\bibfnamefont {H.}~\bibnamefont {Iwamura}},\
  }\bibfield  {title} {\enquote {\bibinfo {title} {Ultrafast 1.55 $\mu$m
  all-optical switching using low-temperature-grown multiple quantum wells},}\
  }\href@noop {} {\bibfield  {journal} {\bibinfo  {journal} {Applied physics
  letters}\ }\textbf {\bibinfo {volume} {68}},\ \bibinfo {pages} {153--155}
  (\bibinfo {year} {1996})}\BibitemShut {NoStop}%
\bibitem [{\citenamefont {Prucnal}\ \emph {et~al.}(2016)\citenamefont
  {Prucnal}, \citenamefont {Shastri}, \citenamefont {de~Lima}, \citenamefont
  {Nahmias},\ and\ \citenamefont {Tait}}]{Prucnal:16}%
  \BibitemOpen
  \bibfield  {author} {\bibinfo {author} {\bibfnamefont {P.~R.}\ \bibnamefont
  {Prucnal}}, \bibinfo {author} {\bibfnamefont {B.~J.}\ \bibnamefont
  {Shastri}}, \bibinfo {author} {\bibfnamefont {T.~F.}\ \bibnamefont
  {de~Lima}}, \bibinfo {author} {\bibfnamefont {M.~A.}\ \bibnamefont
  {Nahmias}}, \ and\ \bibinfo {author} {\bibfnamefont {A.~N.}\ \bibnamefont
  {Tait}},\ }\bibfield  {title} {\enquote {\bibinfo {title} {Recent progress in
  semiconductor excitable lasers for photonic spike processing},}\ }\href
  {\doibase 10.1364/AOP.8.000228} {\bibfield  {journal} {\bibinfo  {journal}
  {Advances in Optics and Photonics}\ }\textbf {\bibinfo {volume} {8}},\
  \bibinfo {pages} {228--299} (\bibinfo {year} {2016})}\BibitemShut {NoStop}%
\bibitem [{\citenamefont {Feldmann}\ \emph {et~al.}(2019)\citenamefont
  {Feldmann}, \citenamefont {Youngblood}, \citenamefont {Wright}, \citenamefont
  {Bhaskaran},\ and\ \citenamefont {Pernice}}]{Feldmann:2019aa}%
  \BibitemOpen
  \bibfield  {author} {\bibinfo {author} {\bibfnamefont {J.}~\bibnamefont
  {Feldmann}}, \bibinfo {author} {\bibfnamefont {N.}~\bibnamefont
  {Youngblood}}, \bibinfo {author} {\bibfnamefont {C.~D.}\ \bibnamefont
  {Wright}}, \bibinfo {author} {\bibfnamefont {H.}~\bibnamefont {Bhaskaran}}, \
  and\ \bibinfo {author} {\bibfnamefont {W.~H.~P.}\ \bibnamefont {Pernice}},\
  }\bibfield  {title} {\enquote {\bibinfo {title} {All-optical spiking
  neurosynaptic networks with self-learning capabilities},}\ }\href {\doibase
  10.1038/s41586-019-1157-8} {\bibfield  {journal} {\bibinfo  {journal}
  {Nature}\ }\textbf {\bibinfo {volume} {569}},\ \bibinfo {pages} {208--214}
  (\bibinfo {year} {2019})}\BibitemShut {NoStop}%
\bibitem [{\citenamefont {Peng}\ \emph
  {et~al.}(2018{\natexlab{a}})\citenamefont {Peng}, \citenamefont {Nahmias},
  \citenamefont {de~Lima}, \citenamefont {Tait},\ and\ \citenamefont
  {Shastri}}]{Peng:2018aa}%
  \BibitemOpen
  \bibfield  {author} {\bibinfo {author} {\bibfnamefont {H.}~\bibnamefont
  {Peng}}, \bibinfo {author} {\bibfnamefont {M.~A.}\ \bibnamefont {Nahmias}},
  \bibinfo {author} {\bibfnamefont {T.~F.}\ \bibnamefont {de~Lima}}, \bibinfo
  {author} {\bibfnamefont {A.~N.}\ \bibnamefont {Tait}}, \ and\ \bibinfo
  {author} {\bibfnamefont {B.~J.}\ \bibnamefont {Shastri}},\ }\bibfield
  {title} {\enquote {\bibinfo {title} {Neuromorphic photonic integrated
  circuits},}\ }\bibfield  {booktitle} {\emph {\bibinfo {booktitle} {IEEE
  Journal of Selected Topics in Quantum Electronics}},\ }\href {\doibase
  10.1109/JSTQE.2018.2840448} {\bibfield  {journal} {\bibinfo  {journal} {IEEE
  Journal of Selected Topics in Quantum Electronics}\ }\textbf {\bibinfo
  {volume} {24}},\ \bibinfo {pages} {1--15} (\bibinfo {year}
  {2018}{\natexlab{a}})}\BibitemShut {NoStop}%
\bibitem [{\citenamefont {Peng}\ \emph {et~al.}(2019)\citenamefont {Peng},
  \citenamefont {Angelatos}, \citenamefont {de~Lima}, \citenamefont {Nahmias},
  \citenamefont {Tait}, \citenamefont {Abbaslou}, \citenamefont {Shastri},\
  and\ \citenamefont {Prucnal}}]{Peng:2019aa}%
  \BibitemOpen
  \bibfield  {author} {\bibinfo {author} {\bibfnamefont {H.}~\bibnamefont
  {Peng}}, \bibinfo {author} {\bibfnamefont {G.}~\bibnamefont {Angelatos}},
  \bibinfo {author} {\bibfnamefont {T.~F.}\ \bibnamefont {de~Lima}}, \bibinfo
  {author} {\bibfnamefont {M.~A.}\ \bibnamefont {Nahmias}}, \bibinfo {author}
  {\bibfnamefont {A.}~\bibnamefont {Tait}}, \bibinfo {author} {\bibfnamefont
  {S.}~\bibnamefont {Abbaslou}}, \bibinfo {author} {\bibfnamefont {B.~J.}\
  \bibnamefont {Shastri}}, \ and\ \bibinfo {author} {\bibfnamefont
  {P.}~\bibnamefont {Prucnal}},\ }\bibfield  {title} {\enquote {\bibinfo
  {title} {Temporal information processing with an integrated laser neuron},}\
  }\bibfield  {booktitle} {\emph {\bibinfo {booktitle} {IEEE Journal of
  Selected Topics in Quantum Electronics}},\ }\href {\doibase
  10.1109/JSTQE.2019.2927582} {\bibfield  {journal} {\bibinfo  {journal} {IEEE
  Journal of Selected Topics in Quantum Electronics}\ ,\ \bibinfo {pages}
  {1--1}} (\bibinfo {year} {2019})}\BibitemShut {NoStop}%
\bibitem [{\citenamefont {Dubbeldam}\ and\ \citenamefont
  {Krauskopf}(1999)}]{Dubbeldam1999}%
  \BibitemOpen
  \bibfield  {author} {\bibinfo {author} {\bibfnamefont {J.~L.~A.}\
  \bibnamefont {Dubbeldam}}\ and\ \bibinfo {author} {\bibfnamefont
  {B.}~\bibnamefont {Krauskopf}},\ }\bibfield  {title} {\enquote {\bibinfo
  {title} {{Self-pulsations of lasers with saturable absorber: dynamics and
  bifurcations}},}\ }\href {\doibase 10.1016/S0030-4018(98)00568-9} {\bibfield
  {journal} {\bibinfo  {journal} {Optics Communications}\ }\textbf {\bibinfo
  {volume} {159}},\ \bibinfo {pages} {325--338} (\bibinfo {year}
  {1999})}\BibitemShut {NoStop}%
\bibitem [{\citenamefont {Nahmias}\ \emph {et~al.}(2013)\citenamefont
  {Nahmias}, \citenamefont {Shastri}, \citenamefont {Tait},\ and\ \citenamefont
  {Prucnal}}]{Nahmias2013}%
  \BibitemOpen
  \bibfield  {author} {\bibinfo {author} {\bibfnamefont {M.~A.}\ \bibnamefont
  {Nahmias}}, \bibinfo {author} {\bibfnamefont {B.~J.}\ \bibnamefont
  {Shastri}}, \bibinfo {author} {\bibfnamefont {A.~N.}\ \bibnamefont {Tait}}, \
  and\ \bibinfo {author} {\bibfnamefont {P.~R.}\ \bibnamefont {Prucnal}},\
  }\bibfield  {title} {\enquote {\bibinfo {title} {{A Leaky Integrate-and-Fire
  Laser Neuron for Ultrafast Cognitive Computing}},}\ }\href
  {http://dx.doi.org/10.1109/JSTQE.2013.2257700} {\bibfield  {journal}
  {\bibinfo  {journal} {IEEE Journal of Selected Topics in Quantum
  Electronics}\ }\textbf {\bibinfo {volume} {19}} (\bibinfo {year}
  {2013})}\BibitemShut {NoStop}%
\bibitem [{\citenamefont {Koch}(1998)}]{Koch:1998}%
  \BibitemOpen
  \bibfield  {author} {\bibinfo {author} {\bibfnamefont {C.}~\bibnamefont
  {Koch}},\ }\href@noop {} {\emph {\bibinfo {title} {Biophysics of Computation:
  Information Processing in Single Neurons (Computational Neuroscience)}}}\
  (\bibinfo  {publisher} {Oxford University Press},\ \bibinfo {year}
  {1998})\BibitemShut {NoStop}%
\bibitem [{\citenamefont {Maass}\ and\ \citenamefont
  {Bishop}(2001)}]{maass2001pulsed}%
  \BibitemOpen
  \bibfield  {author} {\bibinfo {author} {\bibfnamefont {W.}~\bibnamefont
  {Maass}}\ and\ \bibinfo {author} {\bibfnamefont {C.~M.}\ \bibnamefont
  {Bishop}},\ }\href@noop {} {\emph {\bibinfo {title} {{Pulsed neural
  networks}}}}\ (\bibinfo  {publisher} {MIT press},\ \bibinfo {address}
  {Cambridge, MA, USA},\ \bibinfo {year} {2001})\BibitemShut {NoStop}%
\bibitem [{\citenamefont {Tait}\ \emph {et~al.}(2014)\citenamefont {Tait},
  \citenamefont {Nahmias}, \citenamefont {Shastri},\ and\ \citenamefont
  {Prucnal}}]{Tait:JLT:2014}%
  \BibitemOpen
  \bibfield  {author} {\bibinfo {author} {\bibfnamefont {A.~N.}\ \bibnamefont
  {Tait}}, \bibinfo {author} {\bibfnamefont {M.~A.}\ \bibnamefont {Nahmias}},
  \bibinfo {author} {\bibfnamefont {B.~J.}\ \bibnamefont {Shastri}}, \ and\
  \bibinfo {author} {\bibfnamefont {P.~R.}\ \bibnamefont {Prucnal}},\
  }\bibfield  {title} {\enquote {\bibinfo {title} {Broadcast and weight: An
  integrated network for scalable photonic spike processing},}\ }\href
  {\doibase 10.1109/JLT.2014.2345652} {\bibfield  {journal} {\bibinfo
  {journal} {Journal of Lightwave Technology}\ }\textbf {\bibinfo {volume}
  {32}},\ \bibinfo {pages} {3427--3439} (\bibinfo {year} {2014})}\BibitemShut
  {NoStop}%
\bibitem [{\citenamefont {Tait}\ \emph {et~al.}(2016)\citenamefont {Tait},
  \citenamefont {Wu}, \citenamefont {de~Lima}, \citenamefont {Zhou},
  \citenamefont {Shastri}, \citenamefont {Nahmias},\ and\ \citenamefont
  {Prucnal}}]{Tait:16anal}%
  \BibitemOpen
  \bibfield  {author} {\bibinfo {author} {\bibfnamefont {A.~N.}\ \bibnamefont
  {Tait}}, \bibinfo {author} {\bibfnamefont {A.~X.}\ \bibnamefont {Wu}},
  \bibinfo {author} {\bibfnamefont {T.~F.}\ \bibnamefont {de~Lima}}, \bibinfo
  {author} {\bibfnamefont {E.}~\bibnamefont {Zhou}}, \bibinfo {author}
  {\bibfnamefont {B.~J.}\ \bibnamefont {Shastri}}, \bibinfo {author}
  {\bibfnamefont {M.~A.}\ \bibnamefont {Nahmias}}, \ and\ \bibinfo {author}
  {\bibfnamefont {P.~R.}\ \bibnamefont {Prucnal}},\ }\bibfield  {title}
  {\enquote {\bibinfo {title} {Microring weight banks},}\ }\href {\doibase
  10.1109/JSTQE.2016.2573583} {\bibfield  {journal} {\bibinfo  {journal} {IEEE
  Journal of Selected Topics in Quantum Electronics}\ }\textbf {\bibinfo
  {volume} {22}},\ \bibinfo {pages} {312--325} (\bibinfo {year}
  {2016})}\BibitemShut {NoStop}%
\bibitem [{\citenamefont {Tait}\ \emph {et~al.}(2018)\citenamefont {Tait},
  \citenamefont {Wu}, \citenamefont {de~Lima}, \citenamefont {Nahmias},
  \citenamefont {Shastri},\ and\ \citenamefont {Prucnal}}]{Tait:18}%
  \BibitemOpen
  \bibfield  {author} {\bibinfo {author} {\bibfnamefont {A.~N.}\ \bibnamefont
  {Tait}}, \bibinfo {author} {\bibfnamefont {A.~X.}\ \bibnamefont {Wu}},
  \bibinfo {author} {\bibfnamefont {T.~F.}\ \bibnamefont {de~Lima}}, \bibinfo
  {author} {\bibfnamefont {M.~A.}\ \bibnamefont {Nahmias}}, \bibinfo {author}
  {\bibfnamefont {B.~J.}\ \bibnamefont {Shastri}}, \ and\ \bibinfo {author}
  {\bibfnamefont {P.~R.}\ \bibnamefont {Prucnal}},\ }\bibfield  {title}
  {\enquote {\bibinfo {title} {Two-pole microring weight banks},}\ }\href
  {\doibase 10.1364/OL.43.002276} {\bibfield  {journal} {\bibinfo  {journal}
  {Opt. Lett.}\ }\textbf {\bibinfo {volume} {43}},\ \bibinfo {pages}
  {2276--2279} (\bibinfo {year} {2018})}\BibitemShut {NoStop}%
\bibitem [{\citenamefont {Appeltant}\ \emph {et~al.}(2011)\citenamefont
  {Appeltant}, \citenamefont {Soriano}, \citenamefont {Van~der Sande},
  \citenamefont {Danckaert}, \citenamefont {Massar}, \citenamefont {Dambre},
  \citenamefont {Schrauwen}, \citenamefont {Mirasso},\ and\ \citenamefont
  {Fischer}}]{Appeltant:2011aa}%
  \BibitemOpen
  \bibfield  {author} {\bibinfo {author} {\bibfnamefont {L.}~\bibnamefont
  {Appeltant}}, \bibinfo {author} {\bibfnamefont {M.~C.}\ \bibnamefont
  {Soriano}}, \bibinfo {author} {\bibfnamefont {G.}~\bibnamefont {Van~der
  Sande}}, \bibinfo {author} {\bibfnamefont {J.}~\bibnamefont {Danckaert}},
  \bibinfo {author} {\bibfnamefont {S.}~\bibnamefont {Massar}}, \bibinfo
  {author} {\bibfnamefont {J.}~\bibnamefont {Dambre}}, \bibinfo {author}
  {\bibfnamefont {B.}~\bibnamefont {Schrauwen}}, \bibinfo {author}
  {\bibfnamefont {C.~R.}\ \bibnamefont {Mirasso}}, \ and\ \bibinfo {author}
  {\bibfnamefont {I.}~\bibnamefont {Fischer}},\ }\bibfield  {title} {\enquote
  {\bibinfo {title} {Information processing using a single dynamical node as
  complex system},}\ }\href {http://dx.doi.org/10.1038/ncomms1476} {\bibfield
  {journal} {\bibinfo  {journal} {Nature Communications}\ }\textbf {\bibinfo
  {volume} {2}},\ \bibinfo {pages} {468} (\bibinfo {year} {2011})}\BibitemShut
  {NoStop}%
\bibitem [{\citenamefont {Paquot}\ \emph {et~al.}(2012)\citenamefont {Paquot},
  \citenamefont {Duport}, \citenamefont {Smerieri}, \citenamefont {Dambre},
  \citenamefont {Schrauwen}, \citenamefont {Haelterman},\ and\ \citenamefont
  {Massar}}]{Paquot:2012}%
  \BibitemOpen
  \bibfield  {author} {\bibinfo {author} {\bibfnamefont {Y.}~\bibnamefont
  {Paquot}}, \bibinfo {author} {\bibfnamefont {F.}~\bibnamefont {Duport}},
  \bibinfo {author} {\bibfnamefont {A.}~\bibnamefont {Smerieri}}, \bibinfo
  {author} {\bibfnamefont {J.}~\bibnamefont {Dambre}}, \bibinfo {author}
  {\bibfnamefont {B.}~\bibnamefont {Schrauwen}}, \bibinfo {author}
  {\bibfnamefont {M.}~\bibnamefont {Haelterman}}, \ and\ \bibinfo {author}
  {\bibfnamefont {S.}~\bibnamefont {Massar}},\ }\bibfield  {title} {\enquote
  {\bibinfo {title} {Optoelectronic reservoir computing},}\ }\href
  {http://dx.doi.org/10.1038/srep00287} {\bibfield  {journal} {\bibinfo
  {journal} {Scientific Reports}\ }\textbf {\bibinfo {volume} {2}},\ \bibinfo
  {pages} {287 EP --} (\bibinfo {year} {2012})}\BibitemShut {NoStop}%
\bibitem [{\citenamefont {Brunner}\ \emph {et~al.}(2013)\citenamefont
  {Brunner}, \citenamefont {Soriano}, \citenamefont {Mirasso},\ and\
  \citenamefont {Fischer}}]{Brunner:2013aa}%
  \BibitemOpen
  \bibfield  {author} {\bibinfo {author} {\bibfnamefont {D.}~\bibnamefont
  {Brunner}}, \bibinfo {author} {\bibfnamefont {M.~C.}\ \bibnamefont
  {Soriano}}, \bibinfo {author} {\bibfnamefont {C.~R.}\ \bibnamefont
  {Mirasso}}, \ and\ \bibinfo {author} {\bibfnamefont {I.}~\bibnamefont
  {Fischer}},\ }\bibfield  {title} {\enquote {\bibinfo {title} {Parallel
  photonic information processing at gigabyte per second data rates using
  transient states},}\ }\href {http://dx.doi.org/10.1038/ncomms2368} {\bibfield
   {journal} {\bibinfo  {journal} {Nature Communications}\ }\textbf {\bibinfo
  {volume} {4}},\ \bibinfo {pages} {1364} (\bibinfo {year} {2013})}\BibitemShut
  {NoStop}%
\bibitem [{\citenamefont {Vandoorne}\ \emph {et~al.}(2014)\citenamefont
  {Vandoorne}, \citenamefont {Mechet}, \citenamefont {Van~Vaerenbergh},
  \citenamefont {Fiers}, \citenamefont {Morthier}, \citenamefont {Verstraeten},
  \citenamefont {Schrauwen}, \citenamefont {Dambre},\ and\ \citenamefont
  {Bienstman}}]{Vandoorne:2014aa}%
  \BibitemOpen
  \bibfield  {author} {\bibinfo {author} {\bibfnamefont {K.}~\bibnamefont
  {Vandoorne}}, \bibinfo {author} {\bibfnamefont {P.}~\bibnamefont {Mechet}},
  \bibinfo {author} {\bibfnamefont {T.}~\bibnamefont {Van~Vaerenbergh}},
  \bibinfo {author} {\bibfnamefont {M.}~\bibnamefont {Fiers}}, \bibinfo
  {author} {\bibfnamefont {G.}~\bibnamefont {Morthier}}, \bibinfo {author}
  {\bibfnamefont {D.}~\bibnamefont {Verstraeten}}, \bibinfo {author}
  {\bibfnamefont {B.}~\bibnamefont {Schrauwen}}, \bibinfo {author}
  {\bibfnamefont {J.}~\bibnamefont {Dambre}}, \ and\ \bibinfo {author}
  {\bibfnamefont {P.}~\bibnamefont {Bienstman}},\ }\bibfield  {title} {\enquote
  {\bibinfo {title} {Experimental demonstration of reservoir computing on a
  silicon photonics chip},}\ }\href {http://dx.doi.org/10.1038/ncomms4541}
  {\bibfield  {journal} {\bibinfo  {journal} {Nature Communications}\ }\textbf
  {\bibinfo {volume} {5}},\ \bibinfo {pages} {3541 EP --} (\bibinfo {year}
  {2014})}\BibitemShut {NoStop}%
\bibitem [{\citenamefont {Timurdogan}\ \emph {et~al.}(2014)\citenamefont
  {Timurdogan}, \citenamefont {Sorace-Agaskar}, \citenamefont {Sun},
  \citenamefont {Shah~Hosseini}, \citenamefont {Biberman},\ and\ \citenamefont
  {Watts}}]{Timurdogan:2014aa}%
  \BibitemOpen
  \bibfield  {author} {\bibinfo {author} {\bibfnamefont {E.}~\bibnamefont
  {Timurdogan}}, \bibinfo {author} {\bibfnamefont {C.~M.}\ \bibnamefont
  {Sorace-Agaskar}}, \bibinfo {author} {\bibfnamefont {J.}~\bibnamefont {Sun}},
  \bibinfo {author} {\bibfnamefont {E.}~\bibnamefont {Shah~Hosseini}}, \bibinfo
  {author} {\bibfnamefont {A.}~\bibnamefont {Biberman}}, \ and\ \bibinfo
  {author} {\bibfnamefont {M.~R.}\ \bibnamefont {Watts}},\ }\bibfield  {title}
  {\enquote {\bibinfo {title} {An ultralow power athermal silicon modulator},}\
  }\href {https://doi.org/10.1038/ncomms5008} {\bibfield  {journal} {\bibinfo
  {journal} {Nature Communications}\ }\textbf {\bibinfo {volume} {5}},\
  \bibinfo {pages} {4008 EP --} (\bibinfo {year} {2014})}\BibitemShut {NoStop}%
\bibitem [{\citenamefont {Arakawa}\ and\ \citenamefont
  {Yariv}(1986)}]{Arakawa:1986aa}%
  \BibitemOpen
  \bibfield  {author} {\bibinfo {author} {\bibfnamefont {Y.}~\bibnamefont
  {Arakawa}}\ and\ \bibinfo {author} {\bibfnamefont {A.}~\bibnamefont
  {Yariv}},\ }\bibfield  {title} {\enquote {\bibinfo {title} {Quantum well
  lasers--gain, spectra, dynamics},}\ }\bibfield  {booktitle} {\emph {\bibinfo
  {booktitle} {IEEE Journal of Quantum Electronics}},\ }\href {\doibase
  10.1109/JQE.1986.1073185} {\bibfield  {journal} {\bibinfo  {journal} {IEEE
  Journal of Quantum Electronics}\ }\textbf {\bibinfo {volume} {22}},\ \bibinfo
  {pages} {1887--1899} (\bibinfo {year} {1986})}\BibitemShut {NoStop}%
\bibitem [{\citenamefont {Peng}\ \emph
  {et~al.}(2018{\natexlab{b}})\citenamefont {Peng}, \citenamefont {Nahmias},
  \citenamefont {de~Lima}, \citenamefont {Tait}, \citenamefont {Shastri},\ and\
  \citenamefont {Prucnal}}]{Peng:IPC}%
  \BibitemOpen
  \bibfield  {author} {\bibinfo {author} {\bibfnamefont {H.-T.}\ \bibnamefont
  {Peng}}, \bibinfo {author} {\bibfnamefont {M.~A.}\ \bibnamefont {Nahmias}},
  \bibinfo {author} {\bibfnamefont {T.~F.}\ \bibnamefont {de~Lima}}, \bibinfo
  {author} {\bibfnamefont {A.~N.}\ \bibnamefont {Tait}}, \bibinfo {author}
  {\bibfnamefont {B.~J.}\ \bibnamefont {Shastri}}, \ and\ \bibinfo {author}
  {\bibfnamefont {P.~R.}\ \bibnamefont {Prucnal}},\ }in\ \href {\doibase
  10.1109/IPCon.2018.8527090} {\emph {\bibinfo {booktitle} {2018 IEEE Photonics
  Conference (IPC)}}}\ (\bibinfo {year} {2018})\ pp.\ \bibinfo {pages}
  {1--2}\BibitemShut {NoStop}%
\bibitem [{\citenamefont {Xu}, \citenamefont {Fattal},\ and\ \citenamefont
  {Beausoleil}(2008)}]{Xu:2008aa}%
  \BibitemOpen
  \bibfield  {author} {\bibinfo {author} {\bibfnamefont {Q.}~\bibnamefont
  {Xu}}, \bibinfo {author} {\bibfnamefont {D.}~\bibnamefont {Fattal}}, \ and\
  \bibinfo {author} {\bibfnamefont {R.~G.}\ \bibnamefont {Beausoleil}},\
  }\bibfield  {title} {\enquote {\bibinfo {title} {Silicon microring resonators
  with 1.5-µm radius},}\ }\bibfield  {booktitle} {\emph {\bibinfo {booktitle}
  {Optics Express}},\ }\href {\doibase 10.1364/OE.16.004309} {\bibfield
  {journal} {\bibinfo  {journal} {Optics Express}\ }\textbf {\bibinfo {volume}
  {16}},\ \bibinfo {pages} {4309--4315} (\bibinfo {year} {2008})}\BibitemShut
  {NoStop}%
\bibitem [{\citenamefont {Gabbiani}\ and\ \citenamefont
  {Koch}(1998)}]{Gabbiani:1998}%
  \BibitemOpen
  \bibfield  {author} {\bibinfo {author} {\bibfnamefont {F.}~\bibnamefont
  {Gabbiani}}\ and\ \bibinfo {author} {\bibfnamefont {C.}~\bibnamefont
  {Koch}},\ }\bibfield  {title} {\enquote {\bibinfo {title} {Principles of
  spike train analysis},}\ }\href@noop {} {\bibfield  {journal} {\bibinfo
  {journal} {Methods in neuronal modeling}\ }\textbf {\bibinfo {volume} {12}},\
  \bibinfo {pages} {313--360} (\bibinfo {year} {1998})}\BibitemShut {NoStop}%
\bibitem [{\citenamefont {Takeda}\ \emph {et~al.}(2013)\citenamefont {Takeda},
  \citenamefont {Sato}, \citenamefont {Shinya}, \citenamefont {Nozaki},
  \citenamefont {Kobayashi}, \citenamefont {Taniyama}, \citenamefont {Notomi},
  \citenamefont {Hasebe}, \citenamefont {Kakitsuka},\ and\ \citenamefont
  {Matsuo}}]{Takeda:2013aa}%
  \BibitemOpen
  \bibfield  {author} {\bibinfo {author} {\bibfnamefont {K.}~\bibnamefont
  {Takeda}}, \bibinfo {author} {\bibfnamefont {T.}~\bibnamefont {Sato}},
  \bibinfo {author} {\bibfnamefont {A.}~\bibnamefont {Shinya}}, \bibinfo
  {author} {\bibfnamefont {K.}~\bibnamefont {Nozaki}}, \bibinfo {author}
  {\bibfnamefont {W.}~\bibnamefont {Kobayashi}}, \bibinfo {author}
  {\bibfnamefont {H.}~\bibnamefont {Taniyama}}, \bibinfo {author}
  {\bibfnamefont {M.}~\bibnamefont {Notomi}}, \bibinfo {author} {\bibfnamefont
  {K.}~\bibnamefont {Hasebe}}, \bibinfo {author} {\bibfnamefont
  {T.}~\bibnamefont {Kakitsuka}}, \ and\ \bibinfo {author} {\bibfnamefont
  {S.}~\bibnamefont {Matsuo}},\ }\bibfield  {title} {\enquote {\bibinfo {title}
  {Few-fj/bit data transmissions using directly modulated lambda-scale embedded
  active region photonic-crystal lasers},}\ }\href
  {https://doi.org/10.1038/nphoton.2013.110} {\bibfield  {journal} {\bibinfo
  {journal} {Nature Photonics}\ }\textbf {\bibinfo {volume} {7}},\ \bibinfo
  {pages} {569 EP --} (\bibinfo {year} {2013})}\BibitemShut {NoStop}%
\bibitem [{\citenamefont {Wu}\ \emph {et~al.}(2015)\citenamefont {Wu},
  \citenamefont {Buckley}, \citenamefont {Schaibley}, \citenamefont {Feng},
  \citenamefont {Yan}, \citenamefont {Mandrus}, \citenamefont {Hatami},
  \citenamefont {Yao}, \citenamefont {Vu{\v c}kovi{\'c}}, \citenamefont
  {Majumdar},\ and\ \citenamefont {Xu}}]{Wu:2015aa}%
  \BibitemOpen
  \bibfield  {author} {\bibinfo {author} {\bibfnamefont {S.}~\bibnamefont
  {Wu}}, \bibinfo {author} {\bibfnamefont {S.}~\bibnamefont {Buckley}},
  \bibinfo {author} {\bibfnamefont {J.~R.}\ \bibnamefont {Schaibley}}, \bibinfo
  {author} {\bibfnamefont {L.}~\bibnamefont {Feng}}, \bibinfo {author}
  {\bibfnamefont {J.}~\bibnamefont {Yan}}, \bibinfo {author} {\bibfnamefont
  {D.~G.}\ \bibnamefont {Mandrus}}, \bibinfo {author} {\bibfnamefont
  {F.}~\bibnamefont {Hatami}}, \bibinfo {author} {\bibfnamefont
  {W.}~\bibnamefont {Yao}}, \bibinfo {author} {\bibfnamefont {J.}~\bibnamefont
  {Vu{\v c}kovi{\'c}}}, \bibinfo {author} {\bibfnamefont {A.}~\bibnamefont
  {Majumdar}}, \ and\ \bibinfo {author} {\bibfnamefont {X.}~\bibnamefont
  {Xu}},\ }\bibfield  {title} {\enquote {\bibinfo {title} {Monolayer
  semiconductor nanocavity lasers with ultralow thresholds},}\ }\href
  {https://doi.org/10.1038/nature14290} {\bibfield  {journal} {\bibinfo
  {journal} {Nature}\ }\textbf {\bibinfo {volume} {520}},\ \bibinfo {pages} {69
  EP --} (\bibinfo {year} {2015})}\BibitemShut {NoStop}%
\bibitem [{\citenamefont {Jeong}, \citenamefont {Bae},\ and\ \citenamefont
  {Jeong}(2017)}]{Jeong:2017}%
  \BibitemOpen
  \bibfield  {author} {\bibinfo {author} {\bibfnamefont {G.-S.}\ \bibnamefont
  {Jeong}}, \bibinfo {author} {\bibfnamefont {W.}~\bibnamefont {Bae}}, \ and\
  \bibinfo {author} {\bibfnamefont {D.-K.}\ \bibnamefont {Jeong}},\ }\bibfield
  {title} {\enquote {\bibinfo {title} {Review of cmos integrated circuit
  technologies for high-speed photo-detection},}\ }\href {\doibase
  10.3390/s17091962} {\bibfield  {journal} {\bibinfo  {journal} {Sensors}\
  }\textbf {\bibinfo {volume} {17}} (\bibinfo {year} {2017}),\
  10.3390/s17091962}\BibitemShut {NoStop}%
\bibitem [{\citenamefont {Sun}\ \emph {et~al.}(2015)\citenamefont {Sun},
  \citenamefont {Wade}, \citenamefont {Lee}, \citenamefont {Orcutt},
  \citenamefont {Alloatti}, \citenamefont {Georgas}, \citenamefont {Waterman},
  \citenamefont {Shainline}, \citenamefont {Avizienis}, \citenamefont {Lin},
  \citenamefont {Moss}, \citenamefont {Kumar}, \citenamefont {Pavanello},
  \citenamefont {Atabaki}, \citenamefont {Cook}, \citenamefont {Ou},
  \citenamefont {Leu}, \citenamefont {Chen}, \citenamefont {Asanovi{\'c}},
  \citenamefont {Ram}, \citenamefont {Popovi{\'c}},\ and\ \citenamefont
  {Stojanovi{\'c}}}]{Sun:2015aa}%
  \BibitemOpen
  \bibfield  {author} {\bibinfo {author} {\bibfnamefont {C.}~\bibnamefont
  {Sun}}, \bibinfo {author} {\bibfnamefont {M.~T.}\ \bibnamefont {Wade}},
  \bibinfo {author} {\bibfnamefont {Y.}~\bibnamefont {Lee}}, \bibinfo {author}
  {\bibfnamefont {J.~S.}\ \bibnamefont {Orcutt}}, \bibinfo {author}
  {\bibfnamefont {L.}~\bibnamefont {Alloatti}}, \bibinfo {author}
  {\bibfnamefont {M.~S.}\ \bibnamefont {Georgas}}, \bibinfo {author}
  {\bibfnamefont {A.~S.}\ \bibnamefont {Waterman}}, \bibinfo {author}
  {\bibfnamefont {J.~M.}\ \bibnamefont {Shainline}}, \bibinfo {author}
  {\bibfnamefont {R.~R.}\ \bibnamefont {Avizienis}}, \bibinfo {author}
  {\bibfnamefont {S.}~\bibnamefont {Lin}}, \bibinfo {author} {\bibfnamefont
  {B.~R.}\ \bibnamefont {Moss}}, \bibinfo {author} {\bibfnamefont
  {R.}~\bibnamefont {Kumar}}, \bibinfo {author} {\bibfnamefont
  {F.}~\bibnamefont {Pavanello}}, \bibinfo {author} {\bibfnamefont {A.~H.}\
  \bibnamefont {Atabaki}}, \bibinfo {author} {\bibfnamefont {H.~M.}\
  \bibnamefont {Cook}}, \bibinfo {author} {\bibfnamefont {A.~J.}\ \bibnamefont
  {Ou}}, \bibinfo {author} {\bibfnamefont {J.~C.}\ \bibnamefont {Leu}},
  \bibinfo {author} {\bibfnamefont {Y.-H.}\ \bibnamefont {Chen}}, \bibinfo
  {author} {\bibfnamefont {K.}~\bibnamefont {Asanovi{\'c}}}, \bibinfo {author}
  {\bibfnamefont {R.~J.}\ \bibnamefont {Ram}}, \bibinfo {author} {\bibfnamefont
  {M.}~\bibnamefont {Popovi{\'c}}}, \ and\ \bibinfo {author} {\bibfnamefont
  {V.~M.}\ \bibnamefont {Stojanovi{\'c}}},\ }\bibfield  {title} {\enquote
  {\bibinfo {title} {Single-chip microprocessor that communicates directly
  using light},}\ }\href {https://doi.org/10.1038/nature16454} {\bibfield
  {journal} {\bibinfo  {journal} {Nature}\ }\textbf {\bibinfo {volume} {528}},\
  \bibinfo {pages} {534 EP --} (\bibinfo {year} {2015})}\BibitemShut {NoStop}%
\bibitem [{\citenamefont {Atabaki}\ \emph {et~al.}(2018)\citenamefont
  {Atabaki}, \citenamefont {Moazeni}, \citenamefont {Pavanello}, \citenamefont
  {Gevorgyan}, \citenamefont {Notaros}, \citenamefont {Alloatti}, \citenamefont
  {Wade}, \citenamefont {Sun}, \citenamefont {Kruger}, \citenamefont {Meng},
  \citenamefont {Al~Qubaisi}, \citenamefont {Wang}, \citenamefont {Zhang},
  \citenamefont {Khilo}, \citenamefont {Baiocco}, \citenamefont {Popovi{\'c}},
  \citenamefont {Stojanovi{\'c}},\ and\ \citenamefont {Ram}}]{Atabaki:2018aa}%
  \BibitemOpen
  \bibfield  {author} {\bibinfo {author} {\bibfnamefont {A.~H.}\ \bibnamefont
  {Atabaki}}, \bibinfo {author} {\bibfnamefont {S.}~\bibnamefont {Moazeni}},
  \bibinfo {author} {\bibfnamefont {F.}~\bibnamefont {Pavanello}}, \bibinfo
  {author} {\bibfnamefont {H.}~\bibnamefont {Gevorgyan}}, \bibinfo {author}
  {\bibfnamefont {J.}~\bibnamefont {Notaros}}, \bibinfo {author} {\bibfnamefont
  {L.}~\bibnamefont {Alloatti}}, \bibinfo {author} {\bibfnamefont {M.~T.}\
  \bibnamefont {Wade}}, \bibinfo {author} {\bibfnamefont {C.}~\bibnamefont
  {Sun}}, \bibinfo {author} {\bibfnamefont {S.~A.}\ \bibnamefont {Kruger}},
  \bibinfo {author} {\bibfnamefont {H.}~\bibnamefont {Meng}}, \bibinfo {author}
  {\bibfnamefont {K.}~\bibnamefont {Al~Qubaisi}}, \bibinfo {author}
  {\bibfnamefont {I.}~\bibnamefont {Wang}}, \bibinfo {author} {\bibfnamefont
  {B.}~\bibnamefont {Zhang}}, \bibinfo {author} {\bibfnamefont
  {A.}~\bibnamefont {Khilo}}, \bibinfo {author} {\bibfnamefont {C.~V.}\
  \bibnamefont {Baiocco}}, \bibinfo {author} {\bibfnamefont {M.}~\bibnamefont
  {Popovi{\'c}}}, \bibinfo {author} {\bibfnamefont {V.~M.}\ \bibnamefont
  {Stojanovi{\'c}}}, \ and\ \bibinfo {author} {\bibfnamefont {R.~J.}\
  \bibnamefont {Ram}},\ }\bibfield  {title} {\enquote {\bibinfo {title}
  {Integrating photonics with silicon nanoelectronics for the next generation
  of systems on a chip},}\ }\href {\doibase 10.1038/s41586-018-0028-z}
  {\bibfield  {journal} {\bibinfo  {journal} {Nature}\ }\textbf {\bibinfo
  {volume} {556}},\ \bibinfo {pages} {349--354} (\bibinfo {year}
  {2018})}\BibitemShut {NoStop}%
\bibitem [{\citenamefont {Shastri}\ \emph {et~al.}(2016)\citenamefont
  {Shastri}, \citenamefont {Nahmias}, \citenamefont {Tait}, \citenamefont
  {Rodriguez}, \citenamefont {Wu},\ and\ \citenamefont
  {Prucnal}}]{Shastri:2016aa}%
  \BibitemOpen
  \bibfield  {author} {\bibinfo {author} {\bibfnamefont {B.~J.}\ \bibnamefont
  {Shastri}}, \bibinfo {author} {\bibfnamefont {M.~A.}\ \bibnamefont
  {Nahmias}}, \bibinfo {author} {\bibfnamefont {A.~N.}\ \bibnamefont {Tait}},
  \bibinfo {author} {\bibfnamefont {A.~W.}\ \bibnamefont {Rodriguez}}, \bibinfo
  {author} {\bibfnamefont {B.}~\bibnamefont {Wu}}, \ and\ \bibinfo {author}
  {\bibfnamefont {P.~R.}\ \bibnamefont {Prucnal}},\ }\bibfield  {title}
  {\enquote {\bibinfo {title} {Spike processing with a graphene excitable
  laser},}\ }\href {\doibase 10.1038/srep19126} {\bibfield  {journal} {\bibinfo
   {journal} {Scientific Reports}\ }\textbf {\bibinfo {volume} {6}},\ \bibinfo
  {pages} {19126 EP --} (\bibinfo {year} {2016})}\BibitemShut {NoStop}%
\bibitem [{\citenamefont {K\"{o}ster}\ \emph {et~al.}(2017)\citenamefont
  {K\"{o}ster}, \citenamefont {Webb}, \citenamefont {Wang}, \citenamefont
  {Nassar}, \citenamefont {Bansal}, \citenamefont {Constable}, \citenamefont
  {Elibol}, \citenamefont {Gray}, \citenamefont {Hall}, \citenamefont {Hornof},
  \citenamefont {Khosrowshahi}, \citenamefont {Kloss}, \citenamefont {Pai},\
  and\ \citenamefont {Rao}}]{Koster:2017}%
  \BibitemOpen
  \bibfield  {author} {\bibinfo {author} {\bibfnamefont {U.}~\bibnamefont
  {K\"{o}ster}}, \bibinfo {author} {\bibfnamefont {T.}~\bibnamefont {Webb}},
  \bibinfo {author} {\bibfnamefont {X.}~\bibnamefont {Wang}}, \bibinfo {author}
  {\bibfnamefont {M.}~\bibnamefont {Nassar}}, \bibinfo {author} {\bibfnamefont
  {A.~K.}\ \bibnamefont {Bansal}}, \bibinfo {author} {\bibfnamefont
  {W.}~\bibnamefont {Constable}}, \bibinfo {author} {\bibfnamefont
  {O.}~\bibnamefont {Elibol}}, \bibinfo {author} {\bibfnamefont
  {S.}~\bibnamefont {Gray}}, \bibinfo {author} {\bibfnamefont {S.}~\bibnamefont
  {Hall}}, \bibinfo {author} {\bibfnamefont {L.}~\bibnamefont {Hornof}},
  \bibinfo {author} {\bibfnamefont {A.}~\bibnamefont {Khosrowshahi}}, \bibinfo
  {author} {\bibfnamefont {C.}~\bibnamefont {Kloss}}, \bibinfo {author}
  {\bibfnamefont {R.~J.}\ \bibnamefont {Pai}}, \ and\ \bibinfo {author}
  {\bibfnamefont {N.}~\bibnamefont {Rao}},\ }\bibfield  {title} {\enquote
  {\bibinfo {title} {Flexpoint: An adaptive numerical format for efficient
  training of deep neural networks},}\ }in\ \href
  {http://papers.nips.cc/paper/6771-flexpoint-an-adaptive-numerical-format-for-efficient-training-of-deep-neural-networks.pdf}
  {\emph {\bibinfo {booktitle} {Advances in Neural Information Processing
  Systems 30}}},\ \bibinfo {editor} {edited by\ \bibinfo {editor}
  {\bibfnamefont {I.}~\bibnamefont {Guyon}}, \bibinfo {editor} {\bibfnamefont
  {U.~V.}\ \bibnamefont {Luxburg}}, \bibinfo {editor} {\bibfnamefont
  {S.}~\bibnamefont {Bengio}}, \bibinfo {editor} {\bibfnamefont
  {H.}~\bibnamefont {Wallach}}, \bibinfo {editor} {\bibfnamefont
  {R.}~\bibnamefont {Fergus}}, \bibinfo {editor} {\bibfnamefont
  {S.}~\bibnamefont {Vishwanathan}}, \ and\ \bibinfo {editor} {\bibfnamefont
  {R.}~\bibnamefont {Garnett}}}\ (\bibinfo  {publisher} {Curran Associates,
  Inc.},\ \bibinfo {year} {2017})\ pp.\ \bibinfo {pages}
  {1742--1752}\BibitemShut {NoStop}%
\bibitem [{\citenamefont {Hubara}\ \emph {et~al.}(2016)\citenamefont {Hubara},
  \citenamefont {Courbariaux}, \citenamefont {Soudry}, \citenamefont
  {El-Yaniv},\ and\ \citenamefont {Bengio}}]{Hubara:2016}%
  \BibitemOpen
  \bibfield  {author} {\bibinfo {author} {\bibfnamefont {I.}~\bibnamefont
  {Hubara}}, \bibinfo {author} {\bibfnamefont {M.}~\bibnamefont {Courbariaux}},
  \bibinfo {author} {\bibfnamefont {D.}~\bibnamefont {Soudry}}, \bibinfo
  {author} {\bibfnamefont {R.}~\bibnamefont {El-Yaniv}}, \ and\ \bibinfo
  {author} {\bibfnamefont {Y.}~\bibnamefont {Bengio}},\ }\bibfield  {title}
  {\enquote {\bibinfo {title} {Binarized neural networks},}\ }in\ \href
  {http://papers.nips.cc/paper/6573-binarized-neural-networks.pdf} {\emph
  {\bibinfo {booktitle} {Advances in Neural Information Processing Systems
  29}}},\ \bibinfo {editor} {edited by\ \bibinfo {editor} {\bibfnamefont
  {D.~D.}\ \bibnamefont {Lee}}, \bibinfo {editor} {\bibfnamefont
  {M.}~\bibnamefont {Sugiyama}}, \bibinfo {editor} {\bibfnamefont {U.~V.}\
  \bibnamefont {Luxburg}}, \bibinfo {editor} {\bibfnamefont {I.}~\bibnamefont
  {Guyon}}, \ and\ \bibinfo {editor} {\bibfnamefont {R.}~\bibnamefont
  {Garnett}}}\ (\bibinfo  {publisher} {Curran Associates, Inc.},\ \bibinfo
  {year} {2016})\ pp.\ \bibinfo {pages} {4107--4115}\BibitemShut {NoStop}%
\bibitem [{\citenamefont {Courbariaux}, \citenamefont {Bengio},\ and\
  \citenamefont {David}(2015)}]{Courbariaux:2015}%
  \BibitemOpen
  \bibfield  {author} {\bibinfo {author} {\bibfnamefont {M.}~\bibnamefont
  {Courbariaux}}, \bibinfo {author} {\bibfnamefont {Y.}~\bibnamefont {Bengio}},
  \ and\ \bibinfo {author} {\bibfnamefont {J.-P.}\ \bibnamefont {David}},\
  }\bibfield  {title} {\enquote {\bibinfo {title} {Binaryconnect: Training deep
  neural networks with binary weights during propagations},}\ }in\ \href
  {http://papers.nips.cc/paper/5647-binaryconnect-training-deep-neural-networks-with-binary-weights-during-propagations.pdf}
  {\emph {\bibinfo {booktitle} {Advances in Neural Information Processing
  Systems 28}}},\ \bibinfo {editor} {edited by\ \bibinfo {editor}
  {\bibfnamefont {C.}~\bibnamefont {Cortes}}, \bibinfo {editor} {\bibfnamefont
  {N.~D.}\ \bibnamefont {Lawrence}}, \bibinfo {editor} {\bibfnamefont {D.~D.}\
  \bibnamefont {Lee}}, \bibinfo {editor} {\bibfnamefont {M.}~\bibnamefont
  {Sugiyama}}, \ and\ \bibinfo {editor} {\bibfnamefont {R.}~\bibnamefont
  {Garnett}}}\ (\bibinfo  {publisher} {Curran Associates, Inc.},\ \bibinfo
  {year} {2015})\ pp.\ \bibinfo {pages} {3123--3131}\BibitemShut {NoStop}%
\bibitem [{\citenamefont {Trillo}\ \emph {et~al.}(1988)\citenamefont {Trillo},
  \citenamefont {Wabnitz}, \citenamefont {Wright},\ and\ \citenamefont
  {Stegeman}}]{Trillo:88}%
  \BibitemOpen
  \bibfield  {author} {\bibinfo {author} {\bibfnamefont {S.}~\bibnamefont
  {Trillo}}, \bibinfo {author} {\bibfnamefont {S.}~\bibnamefont {Wabnitz}},
  \bibinfo {author} {\bibfnamefont {E.~M.}\ \bibnamefont {Wright}}, \ and\
  \bibinfo {author} {\bibfnamefont {G.~I.}\ \bibnamefont {Stegeman}},\
  }\bibfield  {title} {\enquote {\bibinfo {title} {Soliton switching in fiber
  nonlinear directional couplers},}\ }\href {\doibase 10.1364/OL.13.000672}
  {\bibfield  {journal} {\bibinfo  {journal} {Opt. Lett.}\ }\textbf {\bibinfo
  {volume} {13}},\ \bibinfo {pages} {672--674} (\bibinfo {year}
  {1988})}\BibitemShut {NoStop}%
\bibitem [{\citenamefont {Nelson}\ \emph {et~al.}(1991)\citenamefont {Nelson},
  \citenamefont {Blow}, \citenamefont {Constantine}, \citenamefont {Doran},
  \citenamefont {Lucek}, \citenamefont {Marshall},\ and\ \citenamefont
  {Smith}}]{Nelson:1991aa}%
  \BibitemOpen
  \bibfield  {author} {\bibinfo {author} {\bibfnamefont {B.~P.}\ \bibnamefont
  {Nelson}}, \bibinfo {author} {\bibfnamefont {K.~J.}\ \bibnamefont {Blow}},
  \bibinfo {author} {\bibfnamefont {P.~D.}\ \bibnamefont {Constantine}},
  \bibinfo {author} {\bibfnamefont {N.~J.}\ \bibnamefont {Doran}}, \bibinfo
  {author} {\bibfnamefont {J.~K.}\ \bibnamefont {Lucek}}, \bibinfo {author}
  {\bibfnamefont {I.~W.}\ \bibnamefont {Marshall}}, \ and\ \bibinfo {author}
  {\bibfnamefont {K.}~\bibnamefont {Smith}},\ }\bibfield  {title} {\enquote
  {\bibinfo {title} {All-optical gbit/s switching using nonlinear optical loop
  mirror},}\ }\bibfield  {booktitle} {\emph {\bibinfo {booktitle} {Electronics
  Letters}},\ }\href {\doibase 10.1049/el:19910438} {\bibfield  {journal}
  {\bibinfo  {journal} {Electronics Letters}\ }\textbf {\bibinfo {volume}
  {27}},\ \bibinfo {pages} {704--705} (\bibinfo {year} {1991})}\BibitemShut
  {NoStop}%
\bibitem [{\citenamefont {Asobe}(1997)}]{Asobe:1997aa}%
  \BibitemOpen
  \bibfield  {author} {\bibinfo {author} {\bibfnamefont {M.}~\bibnamefont
  {Asobe}},\ }\bibfield  {title} {\enquote {\bibinfo {title} {Nonlinear optical
  properties of chalcogenide glass fibers and their application to all-optical
  switching},}\ }\href {\doibase https://doi.org/10.1006/ofte.1997.0214}
  {\bibfield  {journal} {\bibinfo  {journal} {Optical Fiber Technology}\
  }\textbf {\bibinfo {volume} {3}},\ \bibinfo {pages} {142--148} (\bibinfo
  {year} {1997})}\BibitemShut {NoStop}%
\bibitem [{\citenamefont {Tait}(2012)}]{Tait:12}%
  \BibitemOpen
  \bibfield  {author} {\bibinfo {author} {\bibfnamefont {A.~N.}\ \bibnamefont
  {Tait}},\ }\emph {\bibinfo {title} {The Dual Resonator Enhanced Asymmetric
  Mach-Zehnder Interferometer: An Ultrafast Thresholder for Integrated Photonic
  Platforms}},\ \href@noop {} {\bibinfo {type} {Undergraduate thesis}},\
  \bibinfo  {school} {Princeton University} (\bibinfo {year}
  {2012})\BibitemShut {NoStop}%
\bibitem [{\citenamefont {Sokoloff}\ \emph {et~al.}(1993)\citenamefont
  {Sokoloff}, \citenamefont {Prucnal}, \citenamefont {Glesk},\ and\
  \citenamefont {Kane}}]{Sokoloff:1993aa}%
  \BibitemOpen
  \bibfield  {author} {\bibinfo {author} {\bibfnamefont {J.~P.}\ \bibnamefont
  {Sokoloff}}, \bibinfo {author} {\bibfnamefont {P.~R.}\ \bibnamefont
  {Prucnal}}, \bibinfo {author} {\bibfnamefont {I.}~\bibnamefont {Glesk}}, \
  and\ \bibinfo {author} {\bibfnamefont {M.}~\bibnamefont {Kane}},\ }\bibfield
  {title} {\enquote {\bibinfo {title} {A terahertz optical asymmetric
  demultiplexer (toad)},}\ }\bibfield  {booktitle} {\emph {\bibinfo {booktitle}
  {IEEE Photonics Technology Letters}},\ }\href {\doibase 10.1109/68.229807}
  {\bibfield  {journal} {\bibinfo  {journal} {IEEE Photonics Technology
  Letters}\ }\textbf {\bibinfo {volume} {5}},\ \bibinfo {pages} {787--790}
  (\bibinfo {year} {1993})}\BibitemShut {NoStop}%
\bibitem [{\citenamefont {Stubkjaer}(2000)}]{Stubkjaer:2000aa}%
  \BibitemOpen
  \bibfield  {author} {\bibinfo {author} {\bibfnamefont {K.~E.}\ \bibnamefont
  {Stubkjaer}},\ }\bibfield  {title} {\enquote {\bibinfo {title} {Semiconductor
  optical amplifier-based all-optical gates for high-speed optical
  processing},}\ }\bibfield  {booktitle} {\emph {\bibinfo {booktitle} {IEEE
  Journal of Selected Topics in Quantum Electronics}},\ }\href {\doibase
  10.1109/2944.902198} {\bibfield  {journal} {\bibinfo  {journal} {IEEE Journal
  of Selected Topics in Quantum Electronics}\ }\textbf {\bibinfo {volume}
  {6}},\ \bibinfo {pages} {1428--1435} (\bibinfo {year} {2000})}\BibitemShut
  {NoStop}%
\bibitem [{\citenamefont {Vandoorne}\ \emph {et~al.}(2011)\citenamefont
  {Vandoorne}, \citenamefont {Dambre}, \citenamefont {Verstraeten},
  \citenamefont {Schrauwen},\ and\ \citenamefont
  {Bienstman}}]{Vandoorne:2011aa}%
  \BibitemOpen
  \bibfield  {author} {\bibinfo {author} {\bibfnamefont {K.}~\bibnamefont
  {Vandoorne}}, \bibinfo {author} {\bibfnamefont {J.}~\bibnamefont {Dambre}},
  \bibinfo {author} {\bibfnamefont {D.}~\bibnamefont {Verstraeten}}, \bibinfo
  {author} {\bibfnamefont {B.}~\bibnamefont {Schrauwen}}, \ and\ \bibinfo
  {author} {\bibfnamefont {P.}~\bibnamefont {Bienstman}},\ }\bibfield  {title}
  {\enquote {\bibinfo {title} {Parallel reservoir computing using optical
  amplifiers},}\ }\bibfield  {booktitle} {\emph {\bibinfo {booktitle} {IEEE
  Transactions on Neural Networks}},\ }\href {\doibase
  10.1109/TNN.2011.2161771} {\bibfield  {journal} {\bibinfo  {journal} {IEEE
  Transactions on Neural Networks}\ }\textbf {\bibinfo {volume} {22}},\
  \bibinfo {pages} {1469--1481} (\bibinfo {year} {2011})}\BibitemShut {NoStop}%
\bibitem [{\citenamefont {Selmi}\ \emph {et~al.}(2014)\citenamefont {Selmi},
  \citenamefont {Braive}, \citenamefont {Beaudoin}, \citenamefont {Sagnes},
  \citenamefont {Kuszelewicz},\ and\ \citenamefont {Barbay}}]{Selmi:2014}%
  \BibitemOpen
  \bibfield  {author} {\bibinfo {author} {\bibfnamefont {F.}~\bibnamefont
  {Selmi}}, \bibinfo {author} {\bibfnamefont {R.}~\bibnamefont {Braive}},
  \bibinfo {author} {\bibfnamefont {G.}~\bibnamefont {Beaudoin}}, \bibinfo
  {author} {\bibfnamefont {I.}~\bibnamefont {Sagnes}}, \bibinfo {author}
  {\bibfnamefont {R.}~\bibnamefont {Kuszelewicz}}, \ and\ \bibinfo {author}
  {\bibfnamefont {S.}~\bibnamefont {Barbay}},\ }\bibfield  {title} {\enquote
  {\bibinfo {title} {Relative refractory period in an excitable semiconductor
  laser},}\ }\href {\doibase https://doi.org/10.1103/PhysRevLett.112.183902}
  {\bibfield  {journal} {\bibinfo  {journal} {Physical Review Letters}\
  }\textbf {\bibinfo {volume} {112}},\ \bibinfo {pages} {183902} (\bibinfo
  {year} {2014})}\BibitemShut {NoStop}%
\bibitem [{\citenamefont {Romeira}\ \emph {et~al.}(2016)\citenamefont
  {Romeira}, \citenamefont {Av{\'o}}, \citenamefont {Figueiredo}, \citenamefont
  {Barland},\ and\ \citenamefont {Javaloyes}}]{Romeira:2016aa}%
  \BibitemOpen
  \bibfield  {author} {\bibinfo {author} {\bibfnamefont {B.}~\bibnamefont
  {Romeira}}, \bibinfo {author} {\bibfnamefont {R.}~\bibnamefont {Av{\'o}}},
  \bibinfo {author} {\bibfnamefont {J.~L.}\ \bibnamefont {Figueiredo}},
  \bibinfo {author} {\bibfnamefont {S.}~\bibnamefont {Barland}}, \ and\
  \bibinfo {author} {\bibfnamefont {J.}~\bibnamefont {Javaloyes}},\ }\bibfield
  {title} {\enquote {\bibinfo {title} {Regenerative memory in time-delayed
  neuromorphic photonic resonators},}\ }\href
  {http://dx.doi.org/10.1038/srep19510} {\bibfield  {journal} {\bibinfo
  {journal} {Scientific Reports}\ }\textbf {\bibinfo {volume} {6}},\ \bibinfo
  {pages} {19510 EP --} (\bibinfo {year} {2016})}\BibitemShut {NoStop}%
\bibitem [{\citenamefont {Krauskopf}\ \emph {et~al.}(2003)\citenamefont
  {Krauskopf}, \citenamefont {Schneider}, \citenamefont {Sieber}, \citenamefont
  {Wieczorek},\ and\ \citenamefont {Wolfrum}}]{Krauskopf2003}%
  \BibitemOpen
  \bibfield  {author} {\bibinfo {author} {\bibfnamefont {B.}~\bibnamefont
  {Krauskopf}}, \bibinfo {author} {\bibfnamefont {K.}~\bibnamefont
  {Schneider}}, \bibinfo {author} {\bibfnamefont {J.}~\bibnamefont {Sieber}},
  \bibinfo {author} {\bibfnamefont {S.}~\bibnamefont {Wieczorek}}, \ and\
  \bibinfo {author} {\bibfnamefont {M.}~\bibnamefont {Wolfrum}},\ }\bibfield
  {title} {\enquote {\bibinfo {title} {{Excitability and self-pulsations near
  homoclinic bifurcations in semiconductor laser systems}},}\ }\href {\doibase
  10.1016/S0030-4018(02)02239-3} {\bibfield  {journal} {\bibinfo  {journal}
  {Optics Communications}\ }\textbf {\bibinfo {volume} {215}},\ \bibinfo
  {pages} {367--379} (\bibinfo {year} {2003})}\BibitemShut {NoStop}%
\bibitem [{\citenamefont {Nahmias}\ \emph {et~al.}(2016)\citenamefont
  {Nahmias}, \citenamefont {Tait}, \citenamefont {Tolias}, \citenamefont
  {Chang}, \citenamefont {Ferreira~de Lima}, \citenamefont {Shastri},\ and\
  \citenamefont {Prucnal}}]{Nahmias:2016}%
  \BibitemOpen
  \bibfield  {author} {\bibinfo {author} {\bibfnamefont {M.~A.}\ \bibnamefont
  {Nahmias}}, \bibinfo {author} {\bibfnamefont {A.~N.}\ \bibnamefont {Tait}},
  \bibinfo {author} {\bibfnamefont {L.}~\bibnamefont {Tolias}}, \bibinfo
  {author} {\bibfnamefont {M.~P.}\ \bibnamefont {Chang}}, \bibinfo {author}
  {\bibfnamefont {T.}~\bibnamefont {Ferreira~de Lima}}, \bibinfo {author}
  {\bibfnamefont {B.~J.}\ \bibnamefont {Shastri}}, \ and\ \bibinfo {author}
  {\bibfnamefont {P.~R.}\ \bibnamefont {Prucnal}},\ }\bibfield  {title}
  {\enquote {\bibinfo {title} {An integrated analog o/e/o link for
  multi-channel laser neurons},}\ }\href {\doibase 10.1063/1.4945368}
  {\bibfield  {journal} {\bibinfo  {journal} {Applied Physics Letters}\
  }\textbf {\bibinfo {volume} {108}},\ \bibinfo {eid} {151106} (\bibinfo {year}
  {2016})}\BibitemShut {NoStop}%
\bibitem [{\citenamefont {Tait}\ \emph {et~al.}(2017)\citenamefont {Tait},
  \citenamefont {de~Lima}, \citenamefont {Zhou}, \citenamefont {Wu},
  \citenamefont {Nahmias}, \citenamefont {Shastri},\ and\ \citenamefont
  {Prucnal}}]{Tait:2017aa}%
  \BibitemOpen
  \bibfield  {author} {\bibinfo {author} {\bibfnamefont {A.~N.}\ \bibnamefont
  {Tait}}, \bibinfo {author} {\bibfnamefont {T.~F.}\ \bibnamefont {de~Lima}},
  \bibinfo {author} {\bibfnamefont {E.}~\bibnamefont {Zhou}}, \bibinfo {author}
  {\bibfnamefont {A.~X.}\ \bibnamefont {Wu}}, \bibinfo {author} {\bibfnamefont
  {M.~A.}\ \bibnamefont {Nahmias}}, \bibinfo {author} {\bibfnamefont {B.~J.}\
  \bibnamefont {Shastri}}, \ and\ \bibinfo {author} {\bibfnamefont {P.~R.}\
  \bibnamefont {Prucnal}},\ }\bibfield  {title} {\enquote {\bibinfo {title}
  {Neuromorphic photonic networks using silicon photonic weight banks},}\
  }\href {\doibase 10.1038/s41598-017-07754-z} {\bibfield  {journal} {\bibinfo
  {journal} {Scientific Reports}\ }\textbf {\bibinfo {volume} {7}},\ \bibinfo
  {pages} {7430} (\bibinfo {year} {2017})}\BibitemShut {NoStop}%
\bibitem [{\citenamefont {Nozaki}\ \emph {et~al.}(2018)\citenamefont {Nozaki},
  \citenamefont {Matsuo}, \citenamefont {Fujii}, \citenamefont {Takeda},
  \citenamefont {Kuramochi}, \citenamefont {Shinya},\ and\ \citenamefont
  {Notomi}}]{Nozaki:18}%
  \BibitemOpen
  \bibfield  {author} {\bibinfo {author} {\bibfnamefont {K.}~\bibnamefont
  {Nozaki}}, \bibinfo {author} {\bibfnamefont {S.}~\bibnamefont {Matsuo}},
  \bibinfo {author} {\bibfnamefont {T.}~\bibnamefont {Fujii}}, \bibinfo
  {author} {\bibfnamefont {K.}~\bibnamefont {Takeda}}, \bibinfo {author}
  {\bibfnamefont {E.}~\bibnamefont {Kuramochi}}, \bibinfo {author}
  {\bibfnamefont {A.}~\bibnamefont {Shinya}}, \ and\ \bibinfo {author}
  {\bibfnamefont {M.}~\bibnamefont {Notomi}},\ }\bibfield  {title} {\enquote
  {\bibinfo {title} {Ultracompact o-e-o converter based on ff-capacitance
  nanophotonic integration},}\ }in\ \href {\doibase
  10.1364/CLEO_SI.2018.SF3A.3} {\emph {\bibinfo {booktitle} {Conference on
  Lasers and Electro-Optics}}}\ (\bibinfo  {publisher} {Optical Society of
  America},\ \bibinfo {year} {2018})\ p.\ \bibinfo {pages} {SF3A.3}\BibitemShut
  {NoStop}%
\bibitem [{\citenamefont {Chen}\ and\ \citenamefont {Lipson}(2009)}]{Chen:09}%
  \BibitemOpen
  \bibfield  {author} {\bibinfo {author} {\bibfnamefont {L.}~\bibnamefont
  {Chen}}\ and\ \bibinfo {author} {\bibfnamefont {M.}~\bibnamefont {Lipson}},\
  }\bibfield  {title} {\enquote {\bibinfo {title} {Ultra-low capacitance and
  high speed germanium photodetectors on silicon},}\ }\href {\doibase
  10.1364/OE.17.007901} {\bibfield  {journal} {\bibinfo  {journal} {Optics
  Express}\ }\textbf {\bibinfo {volume} {17}},\ \bibinfo {pages} {7901--7906}
  (\bibinfo {year} {2009})}\BibitemShut {NoStop}%
\bibitem [{\citenamefont {Michel}, \citenamefont {Liu},\ and\ \citenamefont
  {Kimerling}(2010)}]{Michel:2010aa}%
  \BibitemOpen
  \bibfield  {author} {\bibinfo {author} {\bibfnamefont {J.}~\bibnamefont
  {Michel}}, \bibinfo {author} {\bibfnamefont {J.}~\bibnamefont {Liu}}, \ and\
  \bibinfo {author} {\bibfnamefont {L.~C.}\ \bibnamefont {Kimerling}},\
  }\bibfield  {title} {\enquote {\bibinfo {title} {High-performance ge-on-si
  photodetectors},}\ }\href {https://doi.org/10.1038/nphoton.2010.157}
  {\bibfield  {journal} {\bibinfo  {journal} {Nature Photonics}\ }\textbf
  {\bibinfo {volume} {4}},\ \bibinfo {pages} {527 EP --} (\bibinfo {year}
  {2010})}\BibitemShut {NoStop}%
\bibitem [{\citenamefont {Dong}\ \emph {et~al.}(2009)\citenamefont {Dong},
  \citenamefont {Liao}, \citenamefont {Feng}, \citenamefont {Liang},
  \citenamefont {Zheng}, \citenamefont {Shafiiha}, \citenamefont {Kung},
  \citenamefont {Qian}, \citenamefont {Li}, \citenamefont {Zheng},
  \citenamefont {Krishnamoorthy},\ and\ \citenamefont {Asghari}}]{Dong:09}%
  \BibitemOpen
  \bibfield  {author} {\bibinfo {author} {\bibfnamefont {P.}~\bibnamefont
  {Dong}}, \bibinfo {author} {\bibfnamefont {S.}~\bibnamefont {Liao}}, \bibinfo
  {author} {\bibfnamefont {D.}~\bibnamefont {Feng}}, \bibinfo {author}
  {\bibfnamefont {H.}~\bibnamefont {Liang}}, \bibinfo {author} {\bibfnamefont
  {D.}~\bibnamefont {Zheng}}, \bibinfo {author} {\bibfnamefont
  {R.}~\bibnamefont {Shafiiha}}, \bibinfo {author} {\bibfnamefont {C.-C.}\
  \bibnamefont {Kung}}, \bibinfo {author} {\bibfnamefont {W.}~\bibnamefont
  {Qian}}, \bibinfo {author} {\bibfnamefont {G.}~\bibnamefont {Li}}, \bibinfo
  {author} {\bibfnamefont {X.}~\bibnamefont {Zheng}}, \bibinfo {author}
  {\bibfnamefont {A.~V.}\ \bibnamefont {Krishnamoorthy}}, \ and\ \bibinfo
  {author} {\bibfnamefont {M.}~\bibnamefont {Asghari}},\ }\bibfield  {title}
  {\enquote {\bibinfo {title} {Low vpp, ultralow-energy, compact, high-speed
  silicon electro-optic modulator},}\ }\href {\doibase 10.1364/OE.17.022484}
  {\bibfield  {journal} {\bibinfo  {journal} {Opt. Express}\ }\textbf {\bibinfo
  {volume} {17}},\ \bibinfo {pages} {22484--22490} (\bibinfo {year}
  {2009})}\BibitemShut {NoStop}%
\bibitem [{\citenamefont {Wang}\ \emph {et~al.}(2015)\citenamefont {Wang},
  \citenamefont {Tian}, \citenamefont {Pantouvaki}, \citenamefont {Guo},
  \citenamefont {Absil}, \citenamefont {Van~Campenhout}, \citenamefont
  {Merckling},\ and\ \citenamefont {Van~Thourhout}}]{Wang:2015aa}%
  \BibitemOpen
  \bibfield  {author} {\bibinfo {author} {\bibfnamefont {Z.}~\bibnamefont
  {Wang}}, \bibinfo {author} {\bibfnamefont {B.}~\bibnamefont {Tian}}, \bibinfo
  {author} {\bibfnamefont {M.}~\bibnamefont {Pantouvaki}}, \bibinfo {author}
  {\bibfnamefont {W.}~\bibnamefont {Guo}}, \bibinfo {author} {\bibfnamefont
  {P.}~\bibnamefont {Absil}}, \bibinfo {author} {\bibfnamefont
  {J.}~\bibnamefont {Van~Campenhout}}, \bibinfo {author} {\bibfnamefont
  {C.}~\bibnamefont {Merckling}}, \ and\ \bibinfo {author} {\bibfnamefont
  {D.}~\bibnamefont {Van~Thourhout}},\ }\bibfield  {title} {\enquote {\bibinfo
  {title} {Room-temperature inp distributed feedback laser array directly grown
  on silicon},}\ }\href {https://doi.org/10.1038/nphoton.2015.199} {\bibfield
  {journal} {\bibinfo  {journal} {Nature Photonics}\ }\textbf {\bibinfo
  {volume} {9}},\ \bibinfo {pages} {837 EP --} (\bibinfo {year}
  {2015})}\BibitemShut {NoStop}%
\bibitem [{\citenamefont {Liu}\ \emph {et~al.}(2015)\citenamefont {Liu},
  \citenamefont {Srinivasan}, \citenamefont {Norman}, \citenamefont {Gossard},\
  and\ \citenamefont {Bowers}}]{Liu:15}%
  \BibitemOpen
  \bibfield  {author} {\bibinfo {author} {\bibfnamefont {A.~Y.}\ \bibnamefont
  {Liu}}, \bibinfo {author} {\bibfnamefont {S.}~\bibnamefont {Srinivasan}},
  \bibinfo {author} {\bibfnamefont {J.}~\bibnamefont {Norman}}, \bibinfo
  {author} {\bibfnamefont {A.~C.}\ \bibnamefont {Gossard}}, \ and\ \bibinfo
  {author} {\bibfnamefont {J.~E.}\ \bibnamefont {Bowers}},\ }\bibfield  {title}
  {\enquote {\bibinfo {title} {Quantum dot lasers for silicon photonics
  [invited]},}\ }\href {\doibase 10.1364/PRJ.3.0000B1} {\bibfield  {journal}
  {\bibinfo  {journal} {Photon. Res.}\ }\textbf {\bibinfo {volume} {3}},\
  \bibinfo {pages} {B1--B9} (\bibinfo {year} {2015})}\BibitemShut {NoStop}%
\bibitem [{\citenamefont {Lee}(2003)}]{lee2003design}%
  \BibitemOpen
  \bibfield  {author} {\bibinfo {author} {\bibfnamefont {T.~H.}\ \bibnamefont
  {Lee}},\ }\href {\doibase 10.1017/CBO9780511817281} {\emph {\bibinfo {title}
  {The Design of CMOS Radio-Frequency Integrated Circuits}}},\ \bibinfo
  {edition} {2nd}\ ed.\ (\bibinfo  {publisher} {Cambridge University Press},\
  \bibinfo {year} {2003})\BibitemShut {NoStop}%
\bibitem [{\citenamefont {Smith}, \citenamefont {Blaikie},\ and\ \citenamefont
  {Taylor}(1998)}]{Smith:1998aa}%
  \BibitemOpen
  \bibfield  {author} {\bibinfo {author} {\bibfnamefont {E.~D.~J.}\
  \bibnamefont {Smith}}, \bibinfo {author} {\bibfnamefont {R.~J.}\ \bibnamefont
  {Blaikie}}, \ and\ \bibinfo {author} {\bibfnamefont {D.~P.}\ \bibnamefont
  {Taylor}},\ }\bibfield  {title} {\enquote {\bibinfo {title} {Performance
  enhancement of spectral-amplitude-coding optical cdma using pulse-position
  modulation},}\ }\bibfield  {booktitle} {\emph {\bibinfo {booktitle} {IEEE
  Transactions on Communications}},\ }\href {\doibase 10.1109/26.718559}
  {\bibfield  {journal} {\bibinfo  {journal} {IEEE Transactions on
  Communications}\ }\textbf {\bibinfo {volume} {46}},\ \bibinfo {pages}
  {1176--1185} (\bibinfo {year} {1998})}\BibitemShut {NoStop}%
\bibitem [{\citenamefont {Sushchik}\ \emph {et~al.}(2000)\citenamefont
  {Sushchik}, \citenamefont {Rulkov}, \citenamefont {Larson}, \citenamefont
  {Tsimring}, \citenamefont {Abarbanel}, \citenamefont {Yao},\ and\
  \citenamefont {Volkovskii}}]{Sushchik:2000aa}%
  \BibitemOpen
  \bibfield  {author} {\bibinfo {author} {\bibfnamefont {M.}~\bibnamefont
  {Sushchik}}, \bibinfo {author} {\bibfnamefont {N.}~\bibnamefont {Rulkov}},
  \bibinfo {author} {\bibfnamefont {L.}~\bibnamefont {Larson}}, \bibinfo
  {author} {\bibfnamefont {L.}~\bibnamefont {Tsimring}}, \bibinfo {author}
  {\bibfnamefont {H.}~\bibnamefont {Abarbanel}}, \bibinfo {author}
  {\bibfnamefont {K.}~\bibnamefont {Yao}}, \ and\ \bibinfo {author}
  {\bibfnamefont {A.}~\bibnamefont {Volkovskii}},\ }\bibfield  {title}
  {\enquote {\bibinfo {title} {Chaotic pulse position modulation: a robust
  method of communicating with chaos},}\ }\bibfield  {booktitle} {\emph
  {\bibinfo {booktitle} {IEEE Communications Letters}},\ }\href {\doibase
  10.1109/4234.841319} {\bibfield  {journal} {\bibinfo  {journal} {IEEE
  Communications Letters}\ }\textbf {\bibinfo {volume} {4}},\ \bibinfo {pages}
  {128--130} (\bibinfo {year} {2000})}\BibitemShut {NoStop}%
\bibitem [{\citenamefont {Shiu}\ and\ \citenamefont
  {Kahn}(1999)}]{Shiu:1999aa}%
  \BibitemOpen
  \bibfield  {author} {\bibinfo {author} {\bibfnamefont {D.-S.}\ \bibnamefont
  {Shiu}}\ and\ \bibinfo {author} {\bibfnamefont {J.~M.}\ \bibnamefont
  {Kahn}},\ }\bibfield  {title} {\enquote {\bibinfo {title} {Differential
  pulse-position modulation for power-efficient optical communication},}\
  }\bibfield  {booktitle} {\emph {\bibinfo {booktitle} {IEEE Transactions on
  Communications}},\ }\href {\doibase 10.1109/26.780456} {\bibfield  {journal}
  {\bibinfo  {journal} {IEEE Transactions on Communications}\ }\textbf
  {\bibinfo {volume} {47}},\ \bibinfo {pages} {1201--1210} (\bibinfo {year}
  {1999})}\BibitemShut {NoStop}%
\bibitem [{\citenamefont {Merolla}\ \emph {et~al.}(2014)\citenamefont
  {Merolla}, \citenamefont {Arthur}, \citenamefont {Alvarez-Icaza},
  \citenamefont {Cassidy}, \citenamefont {Sawada}, \citenamefont {Akopyan},
  \citenamefont {Jackson}, \citenamefont {Imam}, \citenamefont {Guo},
  \citenamefont {Nakamura}, \citenamefont {Brezzo}, \citenamefont {Vo},
  \citenamefont {Esser}, \citenamefont {Appuswamy}, \citenamefont {Taba},
  \citenamefont {Amir}, \citenamefont {Flickner}, \citenamefont {Risk},
  \citenamefont {Manohar},\ and\ \citenamefont {Modha}}]{Merolla668}%
  \BibitemOpen
  \bibfield  {author} {\bibinfo {author} {\bibfnamefont {P.~A.}\ \bibnamefont
  {Merolla}}, \bibinfo {author} {\bibfnamefont {J.~V.}\ \bibnamefont {Arthur}},
  \bibinfo {author} {\bibfnamefont {R.}~\bibnamefont {Alvarez-Icaza}}, \bibinfo
  {author} {\bibfnamefont {A.~S.}\ \bibnamefont {Cassidy}}, \bibinfo {author}
  {\bibfnamefont {J.}~\bibnamefont {Sawada}}, \bibinfo {author} {\bibfnamefont
  {F.}~\bibnamefont {Akopyan}}, \bibinfo {author} {\bibfnamefont {B.~L.}\
  \bibnamefont {Jackson}}, \bibinfo {author} {\bibfnamefont {N.}~\bibnamefont
  {Imam}}, \bibinfo {author} {\bibfnamefont {C.}~\bibnamefont {Guo}}, \bibinfo
  {author} {\bibfnamefont {Y.}~\bibnamefont {Nakamura}}, \bibinfo {author}
  {\bibfnamefont {B.}~\bibnamefont {Brezzo}}, \bibinfo {author} {\bibfnamefont
  {I.}~\bibnamefont {Vo}}, \bibinfo {author} {\bibfnamefont {S.~K.}\
  \bibnamefont {Esser}}, \bibinfo {author} {\bibfnamefont {R.}~\bibnamefont
  {Appuswamy}}, \bibinfo {author} {\bibfnamefont {B.}~\bibnamefont {Taba}},
  \bibinfo {author} {\bibfnamefont {A.}~\bibnamefont {Amir}}, \bibinfo {author}
  {\bibfnamefont {M.~D.}\ \bibnamefont {Flickner}}, \bibinfo {author}
  {\bibfnamefont {W.~P.}\ \bibnamefont {Risk}}, \bibinfo {author}
  {\bibfnamefont {R.}~\bibnamefont {Manohar}}, \ and\ \bibinfo {author}
  {\bibfnamefont {D.~S.}\ \bibnamefont {Modha}},\ }\bibfield  {title} {\enquote
  {\bibinfo {title} {A million spiking-neuron integrated circuit with a
  scalable communication network and interface},}\ }\href {\doibase
  10.1126/science.1254642} {\bibfield  {journal} {\bibinfo  {journal}
  {Science}\ }\textbf {\bibinfo {volume} {345}},\ \bibinfo {pages} {668--673}
  (\bibinfo {year} {2014})},\ \Eprint
  {http://arxiv.org/abs/http://science.sciencemag.org/content/345/6197/668.full.pdf}
  {http://science.sciencemag.org/content/345/6197/668.full.pdf} \BibitemShut
  {NoStop}%
\bibitem [{\citenamefont {Furber}(2016)}]{Furber:2016}%
  \BibitemOpen
  \bibfield  {author} {\bibinfo {author} {\bibfnamefont {S.}~\bibnamefont
  {Furber}},\ }\bibfield  {title} {\enquote {\bibinfo {title} {Large-scale
  neuromorphic computing systems},}\ }\href
  {http://stacks.iop.org/1741-2552/13/i=5/a=051001} {\bibfield  {journal}
  {\bibinfo  {journal} {Journal of Neural Engineering}\ }\textbf {\bibinfo
  {volume} {13}},\ \bibinfo {pages} {051001} (\bibinfo {year}
  {2016})}\BibitemShut {NoStop}%
\bibitem [{\citenamefont {Pfeiffer}\ and\ \citenamefont
  {Pfeil}(2018)}]{Pfeiffer:2018aa}%
  \BibitemOpen
  \bibfield  {author} {\bibinfo {author} {\bibfnamefont {M.}~\bibnamefont
  {Pfeiffer}}\ and\ \bibinfo {author} {\bibfnamefont {T.}~\bibnamefont
  {Pfeil}},\ }\bibfield  {title} {\enquote {\bibinfo {title} {Deep learning
  with spiking neurons: Opportunities and challenges},}\ }\href {\doibase
  10.3389/fnins.2018.00774} {\bibfield  {journal} {\bibinfo  {journal}
  {Frontiers in neuroscience}\ }\textbf {\bibinfo {volume} {12}},\ \bibinfo
  {pages} {774; 774--774} (\bibinfo {year} {2018})}\BibitemShut {NoStop}%
\bibitem [{\citenamefont {Samadi}, \citenamefont {Lillicrap},\ and\
  \citenamefont {Tweed}(2017)}]{Samadi:2017aa}%
  \BibitemOpen
  \bibfield  {author} {\bibinfo {author} {\bibfnamefont {A.}~\bibnamefont
  {Samadi}}, \bibinfo {author} {\bibfnamefont {T.~P.}\ \bibnamefont
  {Lillicrap}}, \ and\ \bibinfo {author} {\bibfnamefont {D.~B.}\ \bibnamefont
  {Tweed}},\ }\bibfield  {title} {\enquote {\bibinfo {title} {Deep learning
  with dynamic spiking neurons and fixed feedback weights},}\ }\bibfield
  {booktitle} {\emph {\bibinfo {booktitle} {Neural Computation}},\ }\href
  {\doibase 10.1162/NECO{\_}a{\_}00929} {\bibfield  {journal} {\bibinfo
  {journal} {Neural Computation}\ }\textbf {\bibinfo {volume} {29}},\ \bibinfo
  {pages} {578--602} (\bibinfo {year} {2017})}\BibitemShut {NoStop}%
\bibitem [{\citenamefont {Kheradpisheh}\ \emph {et~al.}(2018)\citenamefont
  {Kheradpisheh}, \citenamefont {Ganjtabesh}, \citenamefont {Thorpe},\ and\
  \citenamefont {Masquelier}}]{Kheradpisheh:2018}%
  \BibitemOpen
  \bibfield  {author} {\bibinfo {author} {\bibfnamefont {S.~R.}\ \bibnamefont
  {Kheradpisheh}}, \bibinfo {author} {\bibfnamefont {M.}~\bibnamefont
  {Ganjtabesh}}, \bibinfo {author} {\bibfnamefont {S.~J.}\ \bibnamefont
  {Thorpe}}, \ and\ \bibinfo {author} {\bibfnamefont {T.}~\bibnamefont
  {Masquelier}},\ }\bibfield  {title} {\enquote {\bibinfo {title} {Stdp-based
  spiking deep convolutional neural networks for object recognition},}\ }\href
  {\doibase https://doi.org/10.1016/j.neunet.2017.12.005} {\bibfield  {journal}
  {\bibinfo  {journal} {Neural Networks}\ }\textbf {\bibinfo {volume} {99}},\
  \bibinfo {pages} {56 -- 67} (\bibinfo {year} {2018})}\BibitemShut {NoStop}%
\bibitem [{\citenamefont {Tavanaei}\ \emph {et~al.}(2018)\citenamefont
  {Tavanaei}, \citenamefont {Ghodrati}, \citenamefont {Kheradpisheh},
  \citenamefont {Masquelier},\ and\ \citenamefont {Maida}}]{Tavanaei:2018}%
  \BibitemOpen
  \bibfield  {author} {\bibinfo {author} {\bibfnamefont {A.}~\bibnamefont
  {Tavanaei}}, \bibinfo {author} {\bibfnamefont {M.}~\bibnamefont {Ghodrati}},
  \bibinfo {author} {\bibfnamefont {S.~R.}\ \bibnamefont {Kheradpisheh}},
  \bibinfo {author} {\bibfnamefont {T.}~\bibnamefont {Masquelier}}, \ and\
  \bibinfo {author} {\bibfnamefont {A.~S.}\ \bibnamefont {Maida}},\ }\bibfield
  {title} {\enquote {\bibinfo {title} {Deep learning in spiking neural
  networks},}\ }\href@noop {} {\bibfield  {journal} {\bibinfo  {journal} {arXiv
  preprint arXiv:1804.08150}\ } (\bibinfo {year} {2018})}\BibitemShut {NoStop}%
\bibitem [{\citenamefont {Tait}\ \emph {et~al.}(2015)\citenamefont {Tait},
  \citenamefont {Chang}, \citenamefont {Shastri}, \citenamefont {Nahmias},\
  and\ \citenamefont {Prucnal}}]{Tait:15}%
  \BibitemOpen
  \bibfield  {author} {\bibinfo {author} {\bibfnamefont {A.~N.}\ \bibnamefont
  {Tait}}, \bibinfo {author} {\bibfnamefont {J.}~\bibnamefont {Chang}},
  \bibinfo {author} {\bibfnamefont {B.~J.}\ \bibnamefont {Shastri}}, \bibinfo
  {author} {\bibfnamefont {M.~A.}\ \bibnamefont {Nahmias}}, \ and\ \bibinfo
  {author} {\bibfnamefont {P.~R.}\ \bibnamefont {Prucnal}},\ }\bibfield
  {title} {\enquote {\bibinfo {title} {Demonstration of {WDM} weighted addition
  for principal component analysis},}\ }\href {\doibase 10.1364/OE.23.012758}
  {\bibfield  {journal} {\bibinfo  {journal} {Optics Express}\ }\textbf
  {\bibinfo {volume} {23}},\ \bibinfo {pages} {12758--12765} (\bibinfo {year}
  {2015})}\BibitemShut {NoStop}%
\bibitem [{\citenamefont {{Ferreira de Lima}}\ \emph
  {et~al.}(2016)\citenamefont {{Ferreira de Lima}}, \citenamefont {Tait},
  \citenamefont {Nahmias}, \citenamefont {Shastri},\ and\ \citenamefont
  {Prucnal}}]{FerreiradeLima:2015}%
  \BibitemOpen
  \bibfield  {author} {\bibinfo {author} {\bibfnamefont {T.}~\bibnamefont
  {{Ferreira de Lima}}}, \bibinfo {author} {\bibfnamefont {A.~N.}\ \bibnamefont
  {Tait}}, \bibinfo {author} {\bibfnamefont {M.~A.}\ \bibnamefont {Nahmias}},
  \bibinfo {author} {\bibfnamefont {B.~J.}\ \bibnamefont {Shastri}}, \ and\
  \bibinfo {author} {\bibfnamefont {P.~R.}\ \bibnamefont {Prucnal}},\
  }\bibfield  {title} {\enquote {\bibinfo {title} {{Scalable Wideband Principal
  Component Analysis via Microwave Photonics}},}\ }\href@noop {} {\bibfield
  {journal} {\bibinfo  {journal} {IEEE Photonics Journal}\ }\textbf {\bibinfo
  {volume} {In press}} (\bibinfo {year} {2016})}\BibitemShut {NoStop}%
\end{thebibliography}%

\end{document}